\documentclass[aps]{revtex4} 
\usepackage{graphicx}

\def\beq{\begin{equation}}
\def\eeq{\end{equation}}
\def\n{n}
\def\p{p}
\def\d{\delta}
\def\A{{\cal A}}
\def\B{{\cal B}}
\def\C{{\cal C}}
\def\Ha{{H_0}}

\def\H1{{H_1}}
\def\Vs{{V_\p}}
\def\Vn{{V_\n}}
\def\Wn{{W_\n}}
\def\Ws{{W_\p}}
\def\Xn{{X_\n}}
\def\Xs{{X_\p}}
\def\a00{{{\cal A}_0^0}}
\def\b00{{{\cal B}_0^0}}
\def\c00{{{\cal C}_0^0}}
\def\D00{{{\cal D}_0^0}}
\def\d{\delta}
\def\A{{\cal A}}
\def\B{{\cal B}}
\def\C{{\cal C}}

\begin{document}

\voffset 0.5 true in

\title{Oscillations of General Relativistic Superfluid Neutron Stars}

\author{N. Andersson}

\affiliation{Department of Mathematics, University of Southampton, 
Southampton SO17 1BJ, UK}

\author{G. L. Comer}

\affiliation{Department of Physics, Saint Louis University, St.~Louis, 
MO, 63156-0907, USA}

\author{D. Langlois}

\affiliation{Institut d'Astrophysique de Paris, 98bis Boulevard Arago, 
75014 Paris, France}

\date{\today}
	

\begin{abstract}
We develop a general formalism to treat, in general relativity, the 
nonradial oscillations of a superfluid neutron star about static 
(non-rotating) 
configurations. The matter content of these stars
can, as a first approximation, be described by a 
two-fluid model: one fluid is the neutron superfluid, which is believed 
to exist in the core and inner crust of mature neutron stars; the other 
fluid is a conglomerate of all charged constituents (crust nuclei, 
protons, 
electrons, etcetera).  We use a system of equations that governs the 
perturbations both of the metric and of the matter variables, whatever 
the equation of state for the two fluids. The entrainment effect is
explicitly included. We also take the first step towards allowing 
for the superfluid to be confined to a part of the star by allowing for an 
outer envelope composed of ordinary fluid. We derive and implement
the junction conditions 
for the metric and matter variables at the core/envelope 
interface, and briefly
discuss the nature of the involved phase-transition. 
We then determine the frequencies and gravitational-wave
damping times for a simple model equation of state, incorporating
entrainment through an approximation scheme which extends 
present Newtonian results to the general relativistic regime.
We investigate how the quasinormal modes of a superfluid star 
are affected by changes in the entrainment parameter, and unveil 
a series of avoided crossings between the various modes. 
We provide a proof that, unless the equation of state is very special, 
all modes of a two-fluid star must radiate gravitationally.
We also discuss the future detectability of pulsations in a superfluid 
star and argue that it may  be possible (given advances in the 
relevant technology) to use 
gravitational-wave data to constrain the  parameters of
superfluid neutron stars. 
\end{abstract}

\maketitle

\section{Introduction}

Ever since the realisation that Cepheids are stars undergoing radial 
pulsation \cite{ceph}, the oscillation of stars has been an important 
research area.  With much improved sensitivity, observations over the 
last few years have established a plethora of pulsating stars.  The best 
case by far is provided by the Sun for which it is known \cite{sun} that 
many high order pressure p-modes, and perhaps also the gravity g-modes
and the Coriolis restored r-modes, are excited.  By combining 
observational data with theoretical models researchers have been able to 
infer details of the Sun's internal structure, e.g.~the sound speed at 
different depths.  These impressive results of so-called 
``helioseismology'' provide inspiration for further research into 
stellar oscillations and the hope that ``asteroseismology'' \cite{astero} 
will help further unveil details of the structure of distant stars.  The
purpose of this paper 
is to advance our modelling of non-radial oscillations of 
non-rotating, old and cold neutron stars that contain superfluid 
components.

Within the framework of general relativity oscillating compact stars 
provide an interesting potential source of gravitational radiation.  
There are several scenarios in which one would expect a neutron star to 
pulsate wildly, e.g.~following its formation in a gravitational collapse, 
and it is interesting to ask whether the associated gravitational waves 
could be detected on Earth.  Should this be the case, one can hope to use 
the gravitational-wave data to probe the star's interior and possibly put 
constraints on the supranuclear equation of state \cite{astero1,astero2}. 

In the last few years much attention has been focussed on 
gravitational-wave driven mode-instabilities in rapidly rotating stars. 
Ever since the seminal work of Chandrasekhar, Friedman and Schutz 
\cite{C70,F78,FS78} in the 1970s it has been known that gravitational 
waves can drive various modes of oscillation unstable, and recently it 
has been shown that the r-modes are particularly susceptible to the 
instability \cite{A98,FM98}.  Most astrophysical r-mode scenarios regard 
hot young neutron stars, but it has also been proposed \cite{AKS99} that 
the instability may operate in mature accreting neutron stars, e.g.~in 
Low-Mass X-ray binaries.  These stars are expected to have core 
temperatures well below the superfluid transition temperature, and hence 
one would need to account for superfluidity in any detailed r-mode model. 
This is particularly important since mutual friction
may provide a strong dissipative mechanism on any mode in 
a superfluid star \cite{mutual}.
In addition to this, it has been argued that the presence of a viscous
boundary layer at the base of the crust of a mature neutron star will 
lead to a very strong damping on the r-modes \cite{ekman}. 
While the issue of mutual friction
has been discussed by Lindblom and Mendell \cite{LM00}, there are as yet no
studies of dissipation due to 
a ``realistic'' core-crust interface, that is likely to involve
the transition from a regime where superfluid neutrons co-exist
with a lattice of crust nuclei to a region composed of superfluid
neutrons and (possibly superconducting) protons.

Further motivation for our current work is provided by results
from attempts to model the equation of state at supranuclear densities.
Many modern 
equation of state calculations (within relativistic mean field theory) 
predict a sizeable hyperon fraction at high densities. 
However, 
the hyperons act as an effective ``refrigerant'' which means that 
the star would cool very fast. In fact, it has been argued that the 
presence of hyperons would lead to core temperatures well below those 
indicated by observations. The favoured resolution to this problem 
corresponds to the hyperons being superfluid. In addition, neutron stars
may have cores composed of deconfined quarks which may also 
pair into exotic superfluid states. In other words, if we want to 
understand the dynamics of the core of a realistic mature 
neutron star we have to allow for one (or more) partially decoupled
superfluid components. 

The need for an improved understanding of these problems 
was recently emphasized by the suggestion that the 
presence of hyperons in the deep core would lead to a strong bulk 
viscosity wich could potentially completely suppress the unstable r-modes
\cite{pbj,lo01}. This suggestion brings may difficult issues to 
the top of the agenda. For example, a detailed study of
 unstable modes must necessarily account for the fact that a   
superfluid component(s) may move relative to the normal fluid component.  
One might expect this to have a significant effect on the estimates of 
time scales for an unstable mode.  Furthermore, as has been 
pointed out by Andersson and Comer \cite{AC01a}, one would expect 
new classes of r-modes (and indeed other inertial modes) 
to exist in a superfluid star. This means that the superfluid problem is 
likely
to be richer than the ordinary, single fluid, one. Given that
there have not yet been any detailed investigations into these 
issues it may be premature to draw any definite conclusions regarding
the effect that superfluidity may have on mode-instabilities.  

It is clear that a 
considerable amount of work on stars with one, or several, superfluid
components remains to be done before we can claim to have an understanding
of possible instabilities in mature neutron stars. These problems provide 
strong motivation for our current work.  

Studies of oscillating superfluid stars were pioneered by Epstein 
\cite{E88}, who was the first to suggest that there ought to exist modes 
of oscillation that are unique to the superfluid case.  These modes have 
since been calculated both in Newtonian theory \cite{LM94,L95,prix02} and 
general 
relativity \cite{CLL99,AC01b}.  It is now well established that a simple 
two-fluid model of a superfluid neutron star core has two families of 
fluid pulsation modes.  The first of these is essentially the standard 
pressure 
p-modes, for which the two fluids tend to move together.  The superfluid 
modes, on the other hand, are distinguished by the fact that the two fluids
are counter-moving \cite{CLL99,AC01a}.  Another distinguishing 
characteristic of the superfluid modes is that their mode frequencies 
have a fundamental dependence on entrainment.  Andersson and Comer 
\cite{AC01a} have used a local analysis of the Newtonian superfluid 
equations to show that the modes are essentially acoustic in nature, 
with frequencies that depend on the stellar parameters as
\beq
    \omega^2_s \approx {m_\p \over m^*_\p} {l (l + 1) \over r^2} c^2_\p 
                       \ , \label{sfreq}
\eeq
where $c_\p$ is (roughly) the speed of sound in the proton fluid, $m_\p$ 
and $m^*_\p$ are the bare and effective proton masses, respectively, $r$ 
is the radial coordinate, and $l$ is the index of the relevant spherical 
harmonic $Y_{l m}(\theta,\phi)$. These results have recently been 
confirmed by detailed mode-calculations \cite{prix02}. 
The strong dependence of the superfluid modes on 
parameters that are difficult to constrain otherwise 
suggest a strategy that may be used in conjunction with future 
observations to 
narrow down the theoretical uncertainties \cite{AC01b}. 
Specifically, an observational determination of the mode 
frequencies of an oscillating neutron star could be used to constrain 
the proton effective mass in dense nuclear matter.  This would 
have immediate implications for BCS energy gap calculations since the 
effective mass is part of the required information (see, for instance, 
Khodel, Khodel, and Clark \cite{KKC01}).

In this paper we concentrate on the equations that describe polar 
perturbations (even parity) of a general relativistic superfluid neutron 
star, using as our starting point the formalism developed by Comer, 
Langlois, and Lin \cite{CLL99}.  The neutron star is assumed to consist 
of two distinct regions: a core consisting of matter in superfluid 
states, and an  envelope of ordinary fluid matter.  
We report 
progress in three important directions.  Firstly, we have obtained the 
first ever results for gravitational-wave damping rates of the superfluid 
oscillation modes.  This is important as it allows us to assess the 
relevance of these modes for gravitational-wave astronomy.  Secondly, we 
have developed a framework in which a superfluid region can be matched to 
an ordinary fluid region. (Lindblom \& Mendell \cite{LM94} have employed
a similar model in the Newtonian regime.)
The introduction
of an ordinary fluid envelope is the first step towards allowing for the 
presence of
superfluids that are confined to distinct regions in the star. However, 
it is important to point out that
since we do not incorporate elasticity in our model the envelope does not
play the same role as the crust of a mature neutron star. In essence, our 
model
corresponds to a star in which 
the fluid degrees of freedom contain a conglomerate ``proton'' fluid 
(consisting of nuclei and electrons in the envelope, and electrons and 
superconducting protons in the core) that extends from the surface to the 
center of the star and superfluid neutrons that exist only in the 
core.  We work out the
relevant junction/boundary conditions at the core/envelope interface and 
determine the quasinormal modes for such a neutron star model.  Thirdly, 
we have included a simple model for the so-called entrainment effect, 
based on an explicit equation of state and an approximation that assumes 
that the fluid velocities are small compared to the speed of 
light.  This allows us to provide the first detailed results describing 
the effect that entrainment has on the oscillation modes of a superfluid 
star.  As we will show, this leads to the presence of so-called ``avoided 
crossings'' between neighbouring modes as the strength of entrainment 
is varied. 

The main body of the paper begins with a discussion, in Sec.~\ref{eqns}, 
of the basic formalism used to describe  non-radial, linearized 
oscillations of general relativistic superfluid neutron stars.  This 
contains a discussion of the background, spherically symmetric and static 
equilibrium configurations, and the introduction of the variables that are 
used to model the non-radial oscillations.  We also  describe the 
key details of the numerical methods used to solve the linearized 
Einstein/superfluid field equations.  The equation of 
state and  inclusion of the entrainment effect are described in 
Sec.~\ref{aem}, and the results of our numerical analysis are presented 
in Sec.~\ref{numres}.  In Sec.~\ref{nonrad} we prove that all pulsation 
modes
of a superfluid star must radiate gravitational waves, unless the 
equation of state belongs to a very special class. 
 Finally, we consider in Sec.~\ref{gws} the 
question of whether or not superfluid modes, excited, for instance, during 
a pulsar
glitch, can be reliably extracted from gravitational-wave data.  Some 
final remarks are offered in the concluding Sec.~\ref{con}.  For clarity 
of presentation and emphasis of the main physical results, we have 
relegated some of the formal, mathematical details to appendices.  
Appendix I is devoted to a derivation of the junction conditions
used to tie together the core with the envelope.  In 
Appendix II we present an analytical equation of state that can be used 
to incorporate the lowest-order effects of entrainment.  
Finally, in Appendix III a new method is presented which can be used to 
accurately determine  long-lived quasinormal modes.  We will be using
geometrized units (except in Appendix II where the speed of light is 
restored) and ``MTW'' \cite{MTW} conventions throughout the paper.

\section{Perturbations of superfluid stars} \label{eqns}

In this Section we summarise the equations that govern the equilibrium 
configurations of the superfluid core and the normal fluid envelope. 
What must be determined in the core are the neutron and proton number 
densities, and two metric coefficients.  In the envelope only the proton 
number density remains but there are still two metric coefficients to 
specify. We also describe the complicated set of coupled 
linear perturbation equations that need to be integrated, and  
outline the computational strategy that we have adopted. 

In what follows, the center of the star is at radial coordinate 
$r = 0$, the core/envelope interface will be at $r = R_c$, and the 
surface at $r = R$.

\subsection{General Relativistic Superfluid Formalism}

The formalism to be used here, and motivation for it, has been described 
in great detail elsewhere 
\cite{CLL99,C89,CL94,CL95,CL98a,CL98b,LSC98,P00,AC01c}. Hence,  we will
only mention 
the key features here.  The central quantity of the 
superfluid formalism is the so-called master function $\Lambda$.  It 
depends on the three scalars $n^2 = -n_\mu n^\mu$, $p^2 = -p_\mu p^\mu$ 
and $x^2 = -p_\mu n^\mu$ that can be formed from the conserved neutron 
($n^{\mu}$) and proton ($p^{\mu}$) number density currents.   
The master function is such that, when the two fluids are 
co-moving, $-\Lambda(n^2,p^2,x^2)$ corresponds to 
the total thermodynamic energy density.  
Once the master function is 
provided (see Sec.~\ref{numres} for a simple analytic equation of state) 
the stress-energy tensor is given by
\beq
    T^\mu_\nu = \Psi \delta^\mu_\nu + p^\mu\chi_\nu + n^\mu \mu_\nu \ ,
\eeq
where
\beq
    \Psi = \Lambda - n^\rho \mu_\rho - p^\rho \chi_\rho \label{press}
\eeq
is the generalised pressure, and
\begin{eqnarray}
    \mu_\nu &=& {\cal B} n_\nu + {\cal A} p_\nu  \ , \\
    \chi_\nu &=& {\cal A} n_\nu + {\cal C} p_\nu \ , 
\end{eqnarray}
are the chemical potential covectors.  We have also introduced
\beq
    {\cal A} = -{ \partial \Lambda \over \partial x^2} \ , \quad 
    {\cal B} = -2{ \partial \Lambda \over \partial n^2} \ , \quad 
    {\cal C} = -2 { \partial \Lambda \over \partial p^2} \ . \quad 
\eeq
The momentum covectors $\mu_\nu$ and $\chi_\nu$ are dynamically, and 
thermodynamically, conjugate to $n^\nu$ and $\p^\nu$. 
The two covectors also make manifest the so-called entrainment effect,
that affects the dynamics of a superfluid neutron star in a crucial way.
It is easy to see that the momentum of one constituent ($\mu_\nu$, say)
carries along some of the mass current of the other constituent 
if ${\cal A}\neq 0$ (since $\mu_\nu$ 
is a linear combination of $n_\nu$ and $p_\nu$).  On the other hand,
if ${\cal A} = 0$, i.e. if the master function does not depend on $x^2$, 
then
there is no entrainment. 

While the standard one-fluid problem can be expressed solely in terms of 
the 
Einstein equations, with the fluid equations of motion being 
automatically satisfied
``by virtue of the Bianchi identities'', the two-fluid problem 
is 
different. Because of the additional dynamic 
degrees of freedom associated with the second fluid we need to use 
also
(a subset of) the fluid equations of motion.
The equations that need to be solved, in addition to the Einstein field
equations, are two ``continuity'' equations 
\beq
\nabla_\mu n^\mu = \nabla_\mu p^\mu = 0 \ .
\label{conteq}\eeq
These equations represent conservation of the superfluid neutrons
and the protons separately. That is, we ignore ``transfusion'' from
one component to the other due to, for example, weak interactions
\cite{LSC98}. This is likely to be a reasonable approximation for the 
timescales and amplitudes that are 
relevant for nonradial pulsation. We also have
two Euler-type equations
\beq
n^\mu \nabla_{[\mu} \mu _{\nu]} = p^\mu \nabla_{[\mu} \chi _{\nu]} = 0 \ .
\label{euleq}\eeq

\subsection{Equilibrium models}

In order  to determine the background fluid configuration we need to 
evaluate 
the associated metric.  We take our equilibrium configurations to be 
spherically symmetric and static, so the metric can be written in the 
Schwarzschild form
\beq
  ds^2 = - e^{\nu} dt^2 + e^{\lambda}  dr^2 
               + r^2 \left(d\theta^2 + {\rm sin}^2\theta 
                d\phi^2\right) \ . \label{bgmet}
\eeq
The two metric coefficients are determined from two
components of the Einstein equations, 
which can be written in the form
\beq
    \lambda^{\prime} = {1 - e^{\lambda} \over r} - 8 \pi r 
                         e^{\lambda} \Lambda \quad , \quad
    \nu^{\prime} = - {1 - e^{\lambda} \over r} + 8 \pi r 
                         e^{\lambda} \Psi \ , \label{bckgrnd}
\eeq
where a primes represents a radial derivative, and 
it is to be understood that $\Lambda = \Lambda(n,p)$ and 
$\Psi = \Psi(n,p)$ in the interval $0 \leq r \leq R_c$ and $\Lambda = 
\Lambda(p)$ and $\Psi = \Psi(p)$ in the interval $R_c \leq r \leq R$.  

The equations that determine the radial profiles of $\n(r)$ and $\p(r)$ 
in the core have been derived in Comer, Langlois, and Lin \cite{CLL99}.
They follow 
from (\ref{euleq}), and can be written  
\beq
    \A^0_0 \p^{\prime} + \B^0_0 \n^{\prime} + {1 \over 2} (\B \n + A \p) 
           \nu^{\prime} = 0 \quad , \quad
    \C^0_0 \p^{\prime} + \A^0_0 \n^{\prime} + {1 \over 2} (\C \p + \A \n) 
           \nu^{\prime} = 0 \ , \label{bgndfl_c}
\eeq
where
\begin{eqnarray}
\A_0^0 &=& \A + 2 {\partial \B \over \partial \p^2} \n \p + 2 
          {\partial \A \over \partial \n^2} \n^2 + 2 {\partial \A 
          \over \partial \p^2} \p^2 + {\partial \A \over \partial 
          x^2} \p \n \ , \cr
&& \cr
\B_0^0 &=& \B + 2 {\partial \B \over \partial \n^2} \n^2 + 4 
          {\partial \A \over \partial \n^2} \n \p + {\partial \A 
          \over \partial x^2} \p^2 , \cr
&& \cr
\C_0^0 &=& \C + 2 {\partial \C \over\partial \p^2} \p^2 + 4 {\partial 
           \A \over \partial \p^2} \n \p + {\partial A \over \partial 
           x^2} \n^2 \ . \label{coef2}
\end{eqnarray}
In the coefficients above, one sets $x^2 = \n \p$ after the partial 
derivatives are taken for the equilibrium configuration.  

For the particular model
equation of state we use in our numerical calculations, 
the 
above equations require that the two fluids have a common surface
(i.e. $n\to 0$ as $p \to 0$) if one assumes chemical 
equilibrium. The oscillations of such models 
were studied in \cite{CLL99}. Here we want to consider models
such that the two-fluid regime is enclosed within a 
single fluid envelope. It is easy to build such a model by 
relaxing the assumption of chemical equilibrium in the core. 
The resultant models are likely not too unrealistic given the 
relatively long timescales required for nuclear interactions 
to reinstate equilibrium. 

Our models are such that the superfluid neutron number density is zero
in the envelope, and thus the only matter 
equation is
\beq
    \C^0_0 \p^{\prime} + {1 \over 2} C \p \nu^{\prime} = 0 \ , 
    \label{bgndfl_e}
\eeq  
where now
\beq
    \C = - 2 {\partial \Lambda \over \partial \p^2} \quad , \quad
    \c00 = \C + 2 \p^2 {\partial \C \over \partial \p^2} \ .
\eeq

There are three sets of ``boundary'' conditions that must be dealt with: 
a set at the center, one at the interface, and the remaining one at the 
surface of the star.  In view of Eq.~(\ref{bckgrnd}), requiring a 
non-singular behavior at the center of the star will impose that 
$\lambda(0) = 0$, and consequently  $\lambda^{\prime}(0)$ and 
$\nu^{\prime}(0)$ must also vanish.  This in turn implies, in view of 
Eq.~(\ref{bgndfl_c}), that $\p^{\prime}(0)$ and $\n^{\prime}(0)$ have to 
vanish as well. Although our analysis of the junction conditions at
the interface allows for other possibilities, our numerical calculations 
 assume that the phase 
transition that leads to the formation of superfluid neutrons is second 
order. In other words, we consider the  energy density to be continuous at $R_c$.  
From the analysis presented in Appendix~I, we see that the two metric 
coefficients and the pressure must then also be continuous.  We will also 
assume that the proton number density and the proton chemical potential 
are continuous at the interface.  At the surface, we will only consider 
configurations that satisfy $\p(R) = 0$.  A smooth joining of the interior 
spacetime to a Schwarzschild vacuum exterior at the surface of the star 
implies that the total mass $M$ of the system is given by
\beq
    M = - 4 \pi \int^{R}_0 r^2~\Lambda(r) dr
\eeq
and that $\Psi(R) = 0$.  

\subsection{The linearized field equations}

It is well-known that all non-trivial pulsation modes of a nonrotating
fluid  star 
correspond to  polar pertubations (often referred to as ``even parity''). 
In the so-called 
Regge-Wheeler gauge \cite{RW}, the corresponding metric 
components are
\beq
     \delta g_{\mu \nu} = - e^{i \omega t}
            \left[\matrix{e^{\nu}r^l H_{0}(r)&i \omega r^{l+1} 
            H_{1}(r)&0&0 
            \cr i \omega r^{l+1} H_{1}(r)&e^{\lambda} r^l 
            H_{0}(r)&0&0 \cr 
            0&0&r^{l+2} K(r)&0\cr 0&0&0&r^{l+2} {\rm sin}^2\theta K(r)}
            \right]  P_l(\theta) \ . \label{lemet}
\eeq
where $P_l(\theta)$ are the Legendre polynomials.
This decomposition will be applied to both the core and the envelope. 

Writing the neutron and proton four-currents as $n^{\mu} = n 
u^{\mu}$ and $p^{\mu} = p v^{\mu}$, where $u_{\mu} u^{\mu} = - 1$ and 
$v_{\mu} v^{\mu} = - 1$, one can show that in the core the 
independent components are
\beq
    \delta u^i = e^{- \nu/2} {\partial \over \partial t} \delta \xi^i_\n 
                 \quad , \quad
    \delta v^i = e^{- \nu/2} {\partial \over \partial t} \delta \xi^i_\p 
                 \ ,
\eeq
where  
\beq
   \delta \xi_\n^r = e^{-\lambda/2}r^{l - 1}  \Wn(r) P_l 
                      e^{i \omega t} \quad , \quad
   \delta \xi_\n^{\theta} = - r^{l - 2} \Vn(r) 
                {\partial \over \partial \theta} P_l e^{i \omega t} 
                  \ , 
\eeq
and
\beq
   \delta \xi_\p^r = e^{-\lambda/2} r^{l - 1}  \Ws(r) P_l 
                      e^{i \omega t} \quad , \quad
   \delta \xi_\p^{\theta} = - r^{l - 2} \Vs(r) 
                {\partial \over \partial \theta} P_l e^{i \omega t} 
                  \ . 
\eeq
The Lagrangian variations in the neutron and proton number 
densities can be written as, respectively,
\begin{eqnarray}
   {\Delta n \over n} &=& \delta n + n^{\prime} e^{- \lambda/2} 
                                r^{l - 1} W_{\n} \ , \cr
                             && \cr
   {\Delta p \over p} &=& \delta p + p^{\prime} e^{- \lambda/2} 
                                r^{l - 1} W_{\p} \ ,
\end{eqnarray}
and
the conservation equations (\ref{conteq}) for the particle number currents yield
\begin{eqnarray}
   {\Delta n \over n} &=& - r^l \left(e^{- \lambda/2} 
                \left[{l + 1 \over r^2} W_\n + {1 \over r} W^{\prime}_\n
                \right] + {l (l + 1) \over r^2} V_\n - {1 \over 2} H_0 - 
                K\right) \ , \\
                             && \cr
   {\Delta p \over p} &=& - r^l \left(e^{- \lambda/2} 
                \left[{l + 1 \over r^2} W_\p + {1 \over r} W^{\prime}_\p
                \right] + {l (l + 1) \over r^2} V_\p - {1 \over 2} H_0 - 
                K\right) \ . \label{lang}
\end{eqnarray}
Thus, all matter variables can be expressed in terms of the velocity 
variables $W_{\n , \p}$ and $V_{\n , \p}$.  In the envelope, we have 
as the only independent matter variables $\Ws$ and $\Vs$.  

The set of perturbation equations that we solve for in the 
superfluid core have already been listed in \cite{CLL99}, but since
our core/envelope problem requires a slightly different method of 
solution we repeat the relevant equations here. We also need
these equations for the analysis in Section~V.

First we define (in analogy with Lindblom and Detweiler's 
\cite{LD,DL}
approach to the one-fluid problem) 
two new variables as
\beq
    \Xn \equiv \n \left[{e^{\nu/2} \over 2}\mu \Ha + e^{-\nu/2} \omega^2 
    \left(\B \n \Vn + \A \p \Vs\right)\right] - e^{(\nu - \lambda)/2} 
    {\n^{\prime} \over r} \left(\b00 \n \Wn + \a00 \p \Ws\right) \ , 
    \label{xn}
\eeq
and
\beq
    \Xs \equiv \p \left[{e^{\nu/2} \over 2}\chi \Ha + e^{-\nu/2} \omega^2 
    \left(\C \p \Vs + \A \n \Vn\right)\right] - e^{(\nu - \lambda)/2} 
    {\p^{\prime} \over r} \left(\c00 \p \Ws + \a00 \n \Wn\right) \ . 
    \label{xs}
\eeq
Then we find that the Einstein and superfluid field equations yield an
algebraic constraint 
equation
\begin{eqnarray}
  &&e^\lambda \left[{2 - l - l^2 \over r^2} - {3 \over r^2} \left(1 - 
    e^{- \lambda}\right) -8 \pi \Psi\right] \Ha + \left[{2 \omega^2 
    \over e^{\nu}}-{l(l+1)\over 2}e^\lambda\left({1-e^{- \lambda} \over 
    r^2} + 8 \pi \Psi\right)\right] \H1 \cr
  &&
    + \left[-2 e^{\lambda - \nu} \omega^2 + e^\lambda{l^2 + l - 2 \over 
    r^2} + e^{2 \lambda} \left({1 - e^{- \lambda} \over r^2} + 8 \pi \Psi
    \right)\left(1 - {3 \over 2} \left(1 - e^{- \lambda}\right) - 4 \pi 
    r^2 \Psi\right)\right] K \cr
  && 
    + 16 \pi e^{\lambda - \nu/2} \left(\Xn + \Xs\right) = 0 \ , 
    \label{cnstrt}
\end{eqnarray}
 and a system of coupled ordinary 
differential equations (where we use the definition $\D00 = \b00 \c00 - 
(\a00)^2$):
\begin{eqnarray}
    \H1^{\prime} &=& {e^\lambda \over r} \Ha + \left[{\lambda' - \nu' 
    \over 2}-{l+1\over r}\right] \H1 + {e^\lambda \over r} K - 16 
    \pi {e^\lambda \over r} \left(\mu \n \Vn + \chi \p \Vs\right) 
    \ , \label{h1prime} \\
  && \cr
    K^{\prime} &=& {\Ha \over r} + {l (l + 1) \over 2 r} \H1 + 
    \left[{\nu^{\prime} \over 2} - {l + 1 \over r}\right] K - 8 \pi 
    {e^{\lambda/2} \over r}\left[\mu \n \Wn + \chi \p \Ws\right] \ , \\
  && \cr
    \Wn^{\prime} &=& {e^{\lambda/2}r\over 2} \Ha + e^{\lambda/2} r K - 
    e^{\lambda/2} {l (l + 1)\over r} \Vn - \left({l + 1 \over r} + 
    {n^{\prime} \over \n}\right) \Wn + \cr
  &&
    + {\c00 \over \n^2 \D00} \left[e^{(\lambda - \nu)/2} r \Xn + 
    n^{\prime} \left(\b00 \n \Wn + \a00 \p \Ws\right)\right] - \cr
  &&
    {\a00 \over \n \p \D00} \left[e^{(\lambda - \nu)/2}r \Xs + \p^{\prime} 
    \left(\a00 \n \Wn + \c00 \p \Ws\right)\right] \ , \label{Wnprim} \\
  && \cr
    \Ws^{\prime} &=& {e^{\lambda/2} r \over 2} \Ha + e^{\lambda/2} r K - 
    e^{\lambda/2} {l (l + 1) \over r} \Vs - \left({l + 1 \over r} + 
    {\p^{\prime} \over \p}\right) \Ws \cr
  &&
    + {\b00 \over \p^2 \D00} \left[e^{(\lambda - \nu)/2} r \Xs + 
    \p^{\prime} \left(\c00 \p \Ws + \a00 \n \Wn\right)\right] - \cr
  &&
    {\a00\over \n \p \D00} \left[e^{(\lambda - \nu)/2}r \Xn + \n^{\prime} 
    \left(\a00 \p \Ws + \b00 \n \Wn\right)\right] \ , \\
  && \cr
    \Xn^{\prime} &=& - {l \over r} \Xn + {e^{\nu/2} \over 2} \left[\n \mu 
    \left({1 \over r} - \nu^{\prime}\right) - \n^{\prime} \left(\b00 \n 
    + \a00 \p\right)\right] \Ha \cr
  &&
    + \mu \n \left[{e^{\nu/2} \over 4}{l (l + 1) \over r} + {\omega^2 
    \over 2} r e^{-\nu/2}\right] \H1 \cr
  &&
    + e^{\nu/2}\left[\mu \n \left({\nu^{\prime} \over 4} - {1 \over 2 r}
    \right) - \left(\b00 \n + \a00 \p\right) \n^{\prime}\right] K + 
    {l (l + 1) \over r^2} e^{\nu/2} \n^{\prime} \left(\b00 \n \Vn + \a00 
    \p \Vs\right) \cr
  &&
    - e^{(\lambda - \nu)/2} {\omega^2 \over r} \n \left(\B \n \Wn + \A \p 
    \Ws\right) - 4 \pi e^{(\lambda+\nu)/2} {\mu \n \over r} \left(\mu \n 
    \Wn + \chi \p \Ws\right) \cr
  &&
    + e^{-(\lambda - \nu)/2} \left[- {\n^{\prime} \over r} \left(
    \b00^{\prime} \n \Wn + \a00^{\prime} \p \Ws\right) + \left({2 
    \n^{\prime} \over r^2} + {\lambda^{\prime} - \nu^{\prime} \over 2 r} 
    \n^{\prime} - {\n^{\prime \prime} \over r}\right)\right. \cr
  &&
    \left.\left(\b00 \n \Wn + \a00 \p \Ws\right)\right] \ , \\
  && \cr
    \Xs^{\prime} &=& - {l \over r} \Xs + {e^{\nu/2} \over 2} \left[\p \chi
    \left({1 \over r} - \nu^{\prime}\right) - \p^{\prime} \left(\c00 \p + 
    \a00 \n\right)\right] \Ha \cr
  &&
    + \chi \p \left[{e^{\nu/2} \over 4} {l (l + 1) \over r} + {\omega^2 
    \over 2} r e^{-\nu/2}\right] \H1 \cr
  &&
    + e^{\nu/2} \left[\chi \p \left({\nu^{\prime} \over 4} - {1 \over 2 r} 
    \right) - \left(\c00 \p + \a00 \n\right) \p^{\prime}\right] K + {l (l 
    + 1) \over r^2} e^{\nu/2} \p^{\prime} \left(\c00 \p \Vs + \a00 \n 
    \Vn\right) \cr
  &&
    - e^{(\lambda - \nu)/2} {\omega^2 \over r} \p \left(\C \p \Ws + \A \n 
    \Wn\right) - 4 \pi e^{(\lambda + \nu)/2} {\chi \p \over r} \left(\chi 
    \p \Ws + \mu \n \Wn\right) \cr
  &&
    + e^{- (\lambda - \nu)/2} \left[- {\p^{\prime} \over r} \left(
    \c00^{\prime} \p \Ws + \a00^{\prime} \n \Wn\right) + \left({2 
    \p^{\prime} \over r^2} + {\lambda^{\prime} - \nu^{\prime} \over 2 r} 
    \p^{\prime} - {\p^{\prime \prime} \over r}\right)\right.\cr
  &&
    \left.\left(\c00 \p \Ws + \a00 \n \Wn\right)\right] \ . \label{xpprime}
\end{eqnarray}

The equations for the ordinary fluid envelope are obtained by taking
the  $\n = 0$ limit of the field equations. It is, however, not
quite as straightforward as letting  $n\to 0$ in the above set 
of two-fluid perturbation equations. The reason for this is that
we have in places divided through 
by $\D00$ which vanishes in the one-fluid limit.

In the one-fluid case, the   constraint equation becomes
\begin{eqnarray}
   &&e^\lambda \left[{2 - l -l^2 \over r^2} - {3 \over r^2} \left(1 - 
     e^{- \lambda}\right) - 8 \pi \Psi\right] \Ha + \left[{2 \omega^2 
     \over e^{\nu}} - {l (l + 1)\over 2} e^\lambda \left({1 - 
     e^{- \lambda} \over r^2} + 8 \pi \Psi\right)\right] \H1 \cr
   && \cr
   &&+ \left[- 2 e^{\lambda - \nu} \omega^2 + e^\lambda{l^2 + l - 2 \over 
     r^2} + e^{2\lambda} \left({1 - e^{- \lambda} \over r^2} + 8 \pi \Psi
     \right) \left(1 - {3 \over 2} \left(1 - e^{- \lambda}\right) - 4 \pi 
     r^2 \Psi\right)\right]K \cr
   && \cr
   &&+ 16 \pi e^{\lambda - \nu/2} \Xs = 0 \ .
   \label{envcon}
\end{eqnarray}
The other two equations for the metric are
\begin{eqnarray}
   \H1^{\prime} &=& {e^\lambda \over r} \Ha + \left[{\lambda' - \nu' 
                    \over 2} -{l + 1 \over r}\right] \H1 + {e^\lambda 
                    \over r} K - 16 \pi {e^\lambda \over r} \chi \p \Vs 
                    \ , \label{h1env} \cr 
&& \cr
   K^{\prime} &=& {\Ha \over r} + {l (l+1) \over 2 r} \H1 + \left[
                  {\nu^{\prime} \over 2} - {l + 1 \over r}\right] K - 
                  8 \pi {e^{\lambda/2} \over r} \chi \p \Ws \ . 
                  \label{kenv}
\end{eqnarray}
The final two equations are for the fluid and they take the form
\begin{eqnarray}
   \Ws^{\prime} &=& {e^{\lambda/2} r \over 2} \Ha + e^{\lambda/2} r K - 
                    e^{\lambda/2} {l (l + 1) \over r} \Vs - {l + 1 \over 
                    r} \Ws + {e^{(\lambda-\nu)/2} r \over \p^2 \c00} \Xs 
                   \ , \label{wpenv} \cr
                && \cr
   \Xs^{\prime} &=& - {l \over r} \Xs + {e^{\nu/2} \over 2} \left[\p 
                    \chi\left({1 \over r} - \nu^{\prime}\right)
                    - \p^{\prime} \c00 \p\right] \Ha + \chi \p \left[
                    {e^{\nu/2} \over 4}{l (l + 1) \over r} + {\omega^2 
                    \over 2} r e^{-\nu/2}\right] \H1 \cr
                && \cr
                &&+ e^{\nu/2} \left[\chi \p \left({\nu^{\prime} \over 4} 
                  - {1 \over 2 r}\right) - \c00 \p \p^{\prime}\right] K + 
                  {l (l + 1) \over r^2} e^{\nu/2} \p^{\prime} \c00 \p \Vs 
                  \cr
                && \cr
                &&- e^{(\lambda-\nu)/2} {\omega^2 \over r} \C \p^2 
                  \Ws - 4 \pi e^{(\lambda+\nu)/2} {(\chi \p)^2 \over r} 
                  \Ws \cr
                && \cr
                &&+ e^{-(\lambda-\nu)/2} \left[-{\p^{\prime} \over r} 
                  \c00^{\prime} \p \Ws + \left({2 \p^{\prime} 
                  \over r^2} + {\lambda^{\prime}-\nu^{\prime} \over 2 r} 
                  \p^{\prime} - {\p^{\prime \prime} \over r}\right) 
                  \c00 \p \Ws\right] \ , \label{xpenv}
\end{eqnarray}
where
\beq
    \Xs = \p \left[{e^{\nu/2} \over 2}\chi \Ha + e^{-\nu/2} \omega^2 
          \left(\C \p \Vs\right)\right] - e^{(\nu-\lambda)/2} {\p^{\prime} 
          \over r} \left(\c00 \p \Ws\right)  \ .
\eeq
This set of equations is identical to that previously used \cite{DL,LD} 
to study the 
oscillations of normal fluid neutron stars for a range of supranuclear
equations of state.

At the center of the star, the conditions are those given in Appendix A 
of Comer, Langlois, and Lin \cite{CLL99}, i.e. all functions
are regular.  At the surface, the conditions 
are the one-fluid conditions used by Detweiler and Lindblom \cite{DL,LD}.  
The main difference here concerns the interface.  The detailed treatment 
of the interface is discussed in Appendix~I.  It is found that the
relativistic junction conditions imply that the three 
metric perturbations $H_0, H_1$ and $K$  
must be continuous at the interface.  We assume that 
$W_{\n}(R_c)$ is free to vary at the interface, i.e., the
value it takes at the interface is determined by the general 
solution produced for 
the core.  We also assume that $\Xn(R_c) = 0$, which will be shown below 
to be consistent with the chosen equation of state.  From the results 
presented in Appendix~I, it is found that two conditions remain. One of 
these implies that $\Xs$ must be continuous at the interface.  Recalling 
that we assume the proton number density and proton chemical potential to be 
continuous at the interface, i.e. that we have a 
second order phase-transition, the last condition is that $\Ws$ must 
also be continuous at the interface.

\subsection{Computational strategy} \label{cs}

Even though our strategy for integrating the perturbation equations is 
similar to that used by Comer, Langlois, and Lin \cite{CLL99} there are 
some subtleties associated with the presence of the one-fluid envelope.
Hence, it is worthwhile outlining the approach we have taken to the 
problem.  
In the core the perturbation equations can be written as the matrix 
equation
\beq
    {d {\bf Y} \over dr} = {\bf Q} \cdot {\bf Y} \ , 
             \label{ME}
\eeq
where
\beq
     {\bf Y} = \{H_{1},K,W_{\n},W_{\p},X_{\n},X_{\p}\}
\eeq
is an abstract six-dimensional vector field.  The $6 \times 6$ matrix 
${\bf Q}$ depends on $l$, $\omega$, the background fields, and the 
various coefficients $\A$, $\a00$, etc.  As was shown by Comer, Langlois, 
and Lin one need only specify the set of values $\{K(0),\Wn(0),\Ws(0)\}$
at the center of the star.  The remaining variables, $H_1(0)$, $\Xn(0)$, 
and $\Xs(0)$, then follow from the $r\to0$ limit of the 
perturbation equations.  All of the 
second derivatives, $H_1^{\prime \prime}(0)$, $K^{\prime \prime}(0)$, 
etc, are likewise determined by $\{K(0),\Wn(0),\Ws(0)\}$.  This means 
that, in order to provide information that enables us to construct the 
general solution, an integration starting from the centre must generate 
three linearly independent solutions ${\bf Y}_{1}$, ${\bf Y}_{2}$, and 
${\bf Y}_{3}$.  The corresponding general solution can thus be written
\beq
    {\bf Y}(r) = \sum^3_{i = 1} c_{i} {\bf Y}_{i}(r) \ ,
\eeq
where the $c_{i}$ ($i = 1,2,3$) are constants to be determined.

In the envelope, the problem is equivalent to the standard one for a 
single fluid and our strategy is identical to that of Lindblom and 
Detweiler \cite{DL,LD}.  We write the perturbation equations as 
\beq
    {d{\bf \tilde{Y}} \over dr} = {\bf \tilde{Q}} \cdot 
    {\bf \tilde{Y}} \ ,            
\eeq
where
\beq
     {\bf \tilde{Y}} = \{H_{1},K,W_{\p},X_{\p}\}
\eeq
and the matrix ${\bf \tilde{Q}}$ can be deduced from 
Eqns.~(\ref{envcon})--(\ref{xpenv}).  At the surface of the star 
our solution must satisfy the single condition $X_\p(R)=0$ 
(cf.~\cite{DL,LD}).  This means that we must generate three linearly 
independent solutions in the envelope and the general solution can 
therefore be written
\beq
    {\bf\tilde{Y}}(r) = \sum^6_{i = 4} c_{i} {\bf \tilde{Y}}_{i}(r) \ .
\eeq

At the core/envelope interface we must enforce the additional condition 
that $X_{\n}(R_c) = 0$.  We also know, from the analysis in Appendix~I, 
that the variables $H_{1}$, $K$, $W_{\p}$ and $X_{\p}$ should all be 
continuous across the interface.  This means that only the function value 
$W_\n(R_c)$ remains unspecified.  This is, however, as it should be since 
the interface represents a ``free boundary'' for the superfluid neutrons.  
(An analogy can be made with the water/air interface of the oceans, where 
the water oscillation is free at the interface.)  Our strategy 
for solving the problem is based on two steps.  First we continue the 
three solutions from the envelope into the core assuming in addition that 
$W_\n(R_c)=0$.  Then we determine a fourth solution (needed to 
generate the general solution in the core) by assuming that $W_\n(R_c) 
\neq 0$, but that all the other variables vanish at the interface.  
This means 
that we have determined a general solution of form 
\beq
    {\bf Y}(r) = \sum^7_{i = 4} c_{i} {\bf Y}_{i}(r) \ ,
\eeq
where ${\bf Y}_{i}(R_c) ={\bf \tilde{Y}}_i(R_c)$ for $i=4-6$.  

The remaining step is to match the various solutions at some point $r = 
r_m$ in the core $0 \leq r_m \leq R_c$. At $r_m$ we must have
\beq
    \sum^3_{i = 1} c_{i} {\bf Y}_{i}(r_m) = \sum^7_{i = 4} c_{i} 
           {\bf Y}_{i}(r_m) \ . \label{EQ}
\eeq
Once we have provided the overall normalisation (by specifying one of the 
$c_i$ coefficients), this problem can readily be solved for the remaining 
coefficients. This  completes the solution of the interior problem.

To determine the global solution we must also solve for the exterior
metric perturbations. This problem reduces to that of integrating the 
Zerilli equation, cf. \cite{CLL99}. Finally, in order to find a quasinormal
mode (QNM) of the system we need to identify solutions that correspond 
to purely outgoing waves at spatial infinity. Our method for determining 
long-lived QNMs is described in Appendix~III. 
 
\section{An Analytical Entrainment Model} \label{aem}

In order to extend the previous work by Comer, Langlois, and Lin 
\cite{CLL99} further we want to construct a sensible equation of state 
that incorporates the entrainment effect.  This effect arises whenever 
there 
is a coupling between two interpenetrating fluids, and has the net result 
that the momentum of one fluid is not simply proportional to that 
fluid's velocity. Rather, it is a linear combination of the velocities 
of both fluids.  Hence, when one constituent starts to flow it will 
necessarily induce a momentum in the other.  However, the 
entrainment effect is poorly understood, and there are as yet no completed 
relativistic
models that we can base our discussion on. (The problem is currently being considered
by Comer and Joynt, and we are hopeful that a fully relativistic formalism
will soon be available.) The involved 
microphysics is, in fact, so uncertain that the best strategy corresponds 
to introducing some convenient parameterisation and then studying whether 
a variation in the chosen parameters affects the overall properties of the 
star and/or it's modes of pulsation.

We noted earlier, in Sec.~\ref{eqns}, that the two chemical potential 
covectors $\mu_{\mu}$ and $\chi_{\mu}$ make manifest the entrainment 
effect by being linear combinations of the conserved four-currents, 
$\n^{\mu}$ and $\p^{\mu}$.  Ultimately, this  
traces back to the master function $\Lambda$ depending explicitly on the 
``entrainment variable'' $x^2$.  Its presence in the background and 
perturbation equations listed in Sec.~\ref{eqns} is most apparent through 
the values of the coefficients ${\cal A}$ and ${\cal A}_0^0$.  In the 
following we will use an expansion in terms of $x^2$ as outlined in 
Appendix~II.  At the heart of this expansion are the dimensionless 
ratios of the neutron and proton three-velocities with respect to the 
speed of light.  This expansion makes sense since one would expect these 
ratios to be extremely small under most circumstances.  Then the 
expansion becomes particularly accurate.  The equation of state so 
constructed should have applications to the studies of quasinormal modes 
of neutron stars, with both static spherically symmetric and slowly 
rotating backgrounds. We thus expand the master function as
\beq
   \Lambda(n^2,p^2,x^2) = \sum_{i = 0}^{\infty} \lambda_i(n^2,p^2) 
                           \left(x^2 - \n \p\right)^i \ , \label{expand}
\eeq
where $x^2 - \n \p$ is expected to be small with respect to $\n \p$.
It should be noted that there are combinations other than $x^2 - \n \p$ 
that could be used in the expansion, the most obvious being those that 
have no dimensions, say, $(x^2 - \n \p)/n \p$.  These combinations are, 
however,
not convenient in that the $\A$, $\B$, etc coefficients require that 
partial derivatives with respect to $\n^2$ and $\p^2$ be taken.  The 
effect of this is that the expansions for $\A$, $\B$, etc would then not 
take the same form as Eq.~(\ref{expand}) above, since every derivative 
would bring in extra factors of $x^2$ outside the terms $[(x^2 - \n \p)/
\n \p]^i$.  A quick glance at Appendix II will show that if we use 
an expansion based 
on 
$x^2 - \n \p$ the basic form of the expansion is preserved for all  
coefficients that enter the field equations.

In order to make contact with previous Newtonian studies of oscillating 
superfluid stars we will adopt the particular entrainment model used by 
Lindblom and Mendell \cite{LM00}.  To do this, a point of connection must 
be made between the coefficients used in the general relativistic 
superfluid equations and the corresponding
Newtonian equations.  This 
connection can be made by taking the Newtonian limit of the general 
relativistic superfluid formalism and then comparing the dynamical 
variables (as in \cite{AC01a}).  In this way the mass matrix 
components $\rho_{\n \n}$, $\rho_{\p \p}$, and $\rho_{\n \p}$ of the 
Newtonian formalism (cf.~\cite{AB75}) can be directly connected to the 
analytical entrainment coefficient $\lambda_1$ of the general 
relativistic formalism (cf.~Appendix~II) and the particle number 
densities $\n$ and $\p$.  Following this strategy we find that 
$\lambda_1$ can be written as 
\beq
    \lambda_1 = - {c^2 m_{\n} m_{\p} \over \rho_{\n\p}^2 - 
                \rho_{\n\n} \rho_{\p\p}} \rho_{\n\p} \ ,
\eeq
where
\beq
    m_{\n} \n = \rho_{\n \n} + \rho_{\n \p} \quad , \quad
    m_{\p} \p = \rho_{\p \p} + \rho_{\n \p} \ .
\eeq
where $m_{\n(\p)}$ is the neutron (proton) mass.
The particular model of Lindblom and Mendell \cite{LM00} sets
\beq
    \rho_{\n \p} = - \epsilon m_{\n} \n \ ,
\eeq
where $\epsilon$ is a constant.  In the following we will refer to 
$\epsilon$ as the ``entrainment parameter.''  From the analysis of 
Prix, Comer and Andersson \cite{PCA01} one can infer that 
\beq
    \epsilon = {m_\p \p \over m_\n \n} \left({m_\p \over m^*_\p} - 
               1\right) \ ,
\eeq 
where $m^*_\p$ is the 
proton effective mass.  Given that $0.3 \leq m^*_\p/m_\p  \leq 0.8$ (see, 
for instance, Sj\"oberg \cite{S76})  ``typical'' values for a 
neutron star core may lie in the range $0.04 \leq \epsilon \leq 0.2$. 
In the following we take this range as being ``physically reasonable''.

At this point it is relevant to note that, after all the numerical 
work described in this paper had been completed, 
Prix et al \cite{PCA01}  developed an alternative 
description of entrainment. While we could, in principle, have adopted our 
formulas here to this new description we have decided not to do this. 
Such a change would not have affected our discussion or the implications 
of our results in any way. 

\section{Numerical results} \label{numres}

\subsection{An analytical equation of state}

We will now apply our formalism to a simple model equation of state.
We consider the case 
 where each core fluid (at the lowest order in the expansion 
presented in Appendix~II) behaves as a relativistic polytrope, i.e.
we take
\beq
    \lambda_0(\n^2,\p^2) = - m_\n \n - \sigma_{\n} \n^{\beta_{\n}}
                           - m_\p \p - \sigma_{\p} \p^{\beta_{\p}}
                              \ .
\eeq
This master function is clearly separable in $n$ and $p$.  
Using the formalism developed in 
Appendix~II, the  relevant coefficients are (for the equilibrium 
configurations where $x^2 = \n \p$)
\beq
    \A = \epsilon {m_\n m_\p \over m_\p \p + \epsilon (m_\n \n + 
         m_\p \p)} \quad , \quad \a00 = 0 \ , 
\eeq
\beq
    \B = {m_\n \over \n} + \sigma_\n \beta_\n \n^{\beta_\n - 2} - 
         \epsilon {m_\n m_\p \p/\n \over m_\p \p + \epsilon 
         (m_\n \n + m_\p \p)} 
         \quad , \quad 
    \b00 = \sigma_\n \beta_\n \left(\beta_\n - 1\right) \n^{\beta_\n - 2} 
           \ ,
\eeq
and  
\beq
    \C = {m_\p \over \p} + \sigma_\p \beta_\p \p^{\beta_\p - 2} - 
         \epsilon {m_\n m_\p \n/\p \over m_\p \p + \epsilon 
         (m_\n \n + m_\p \p)} 
         \quad , \quad 
    \c00 = \sigma_\p \beta_\p \left(\beta_\p - 1\right)\p^{\beta_\p - 2} 
           \ .
\eeq
The pressure and chemical potentials of the core are
\begin{eqnarray}
    \Psi &=& \sigma_n \left(\beta_n - 1\right) n^{\beta_n} + 
             \sigma_p \left(\beta_p - 1\right) p^{\beta_p} \ , \cr
          && \cr
    \mu &=& m_\n  + \sigma_n \beta_n n^{\beta_n - 1} \ , \cr
         && \cr
    \chi &=& m_\p  + \sigma_p \beta_p p^{\beta_p - 1} \ .
\end{eqnarray}
Furthermore, 
one can verify that these coefficients and thermodynamic variables are 
such that the variable $X_\n$ must vanish at the core/envelope interface, 
cf. Eqn. (\ref{xn}).

Of course, we must also consider what happens in the envelope.  This is 
much simplified because there entrainment is not relevant.  We get
\beq
    \Lambda(\p^2) = - m_\p \p - \sigma_{\p} \p^{\beta_{\p}}
                              \ .
\eeq
The relevant coefficients are
\beq
    \C = {m_\p  \over \p} + \sigma_\p \beta_\p \p^{\beta_\p - 2} 
         \quad , \quad 
    \c00 = \sigma_\p \beta_\p \left(\beta_\p - 1\right)\p^{\beta_\p - 2} 
           \ ,
\eeq
and the pressure and chemical potential are given by
\begin{eqnarray}
    \Psi &=& \sigma_p \left(\beta_p - 1\right) p^{\beta_p} \ , \cr
          && \cr
    \chi &=& m_\p + \sigma_p \beta_p p^{\beta_p - 1} \ .
\end{eqnarray}

\begin{figure}[h]
\centering
\includegraphics[height=6cm,clip]{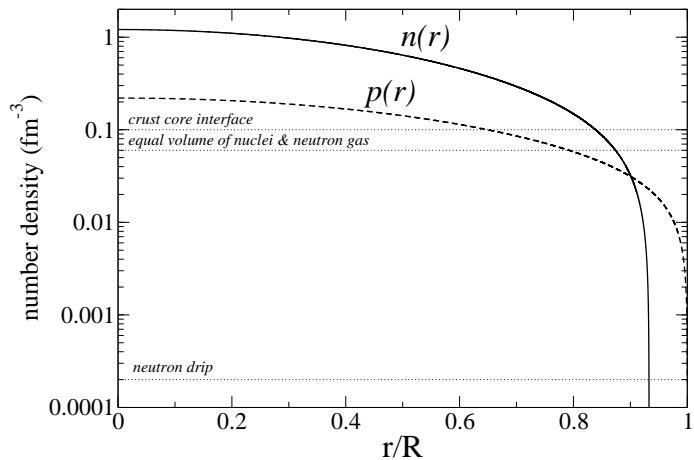}
\caption{The radial profiles of the neutron and proton background 
particle number densities, $n$ and $p$, respectively, for model~II. 
The model has been constructed such that it accords well with a 
$1.4M_\odot$ neutron star determined using the modern equation of state 
calculated by Akmal, Pandharipande and Ravenhall \cite{APR}. 
 For reference, we show as horizontal lines the number densities at 
which Akmal et al suggest that i) neutron drip occurs, ii) there is an 
equal number of nuclei and neutron gas, and iii) the crust/core
interface is located. It should be noted that the latter should not 
coincide with our core/envelope interface since one would expect there 
to be a region where crust nuclei are penetrated by a neutron superfluid. }
\label{backfig}
\end{figure}

We have considered two stellar models.  The first model 
is identical to model~2 
of Comer, Langlois, and Lin \cite{CLL99}, and has no envelope.  
The second model has a significant envelope, and a more realistic 
(slightly smaller) proton 
fraction in the core.  The relevant parameters that determine the 
two models are listed in Table~\ref{modtab}.  In particular, 
the features of the 
core/envelope model are illustrated in Fig.~\ref{backfig}. The figure
shows the radial 
profiles of the background neutron and proton particle number densities, 
$\n(r)$ and $\p(r)$, respectively. The model has been 
constructed  to have the following features that are reckoned to be 
characteristic of real neutron stars: (i) A mass of about $1.4 M_\odot$, 
(ii) a total radius of about $10\ {\rm km}$, (iii) an envelope of roughly 
one kilometer thickness, and (iv) a central proton fraction of about 
$10 \%$.  

\begin{table}
\caption{Parameters describing our stellar models~I and II. Model~I is 
identical to  model~2 of \cite{CLL99}, and has no envelope. 
Model~II, on the other hand, has an envelope of roughly 1~km and could 
be seen as a slightly more realistic neutron star model.}
\begin{tabular}{|c|c|c|}
\hline
 & model I & model II \\
\hline
$\sigma_n/m_n$ & 0.2 & 0.22  \\
$\sigma_p/m_n$ & 0.5 & 1.95  \\
$\beta_n$ & 2.5 & 2.01  \\
$\beta_p$ & 2.0 & 2.38  \\
$n_c$ (fm$^{-3}$) & 1.3 & 1.21  \\
$p_c$ (fm$^{-3}$) & 0.741 & 0.22  \\
$M/M_\odot$ & 1.355 & 1.37   \\
$R\ (\hbox{km})$ & 7.92 &  10.19  \\
$R_{c}\ (\hbox{km})$ & --- & 8.90  \\
\hline
\end{tabular}
\label{modtab}
\end{table}

\subsection{Quasinormal modes}

We have calculated the relevant fluid QNMs for our two stellar models, 
and various values of the entrainment parameter $\epsilon$. To put the 
results in context,  it is useful to recall the basic features of the mode 
spectrum of a non-rotating, ordinary fluid neutron star. Despite  the 
ordinary fluid system being the simplest possible description of the matter 
in a neutron star, there is nevertheless an impressive array of modes.  
In addition to the expected f- and p-modes, which are acoustic in nature, 
there also exists the so-called w-modes \cite{KS92} which are primarily 
due to oscillations of spacetime itself, with little coupling to the 
fluid of the star. If the equation of state has more than one 
parameter, and can be considered stratified --- 
for instance, because of a varying proton fraction ---
then the star can also 
support low-frequency g-modes \cite{RG92} .  

One might expect that the essential doubling of the matter degrees of 
freedom due to the presence of a superfluid component 
might simply lead to a doubling of the families of
modes of oscillation.  However, it should not come as a great
surprise that  the w-modes in a 
superfluid neutron star look very much like those of the ordinary fluid 
case \cite{CLL99}. There is no doubling of modes: The  w-modes are a 
feature of spacetime itself and depend on the curvature 
induced by the background fluid rather than the actual nature of the fluid.  
But it is perhaps surprising that the simple expectation of mode 
doubling is not completely realized for the  modes that are mainly due to 
matter oscillations.  As mentioned in the Introduction, it has long been 
known that the additional fluid degree of freedom leads to the presence of a 
new set of modes, that have been dubbed superfluid 
modes.  They are analogous to the ordinary fluid p-modes in that they are 
predominately acoustic in nature, (cf.~Eq.~(\ref{sfreq})).
The situation regarding g-modes is more confusing. 
Not only are there no new set of pulsating g-modes due 
to superfluidity, the standard set that one might expect to exist 
because of the varying proton fraction do not exist either. 
Inspired by Lee's \cite{L95} numerical results, 
Andersson and Comer \cite{AC01a} have used a 
local analysis of the mode spectrum to show that the g-modes disappear 
from the spectrum of pulsating modes.  They prove that the ``missing'' 
g-modes exist as two independent sets of degenerate modes in 
the space of time-independent 
perturbations (the zero-frequency subspace). Their degeneracy will 
presumably be broken if one were to consider a two-fluid system 
governed by a three parameter equation of state. We plan to discuss
this issue in detail elsewhere.

\begin{figure}[h]
\centering
\includegraphics[height=7cm,clip]{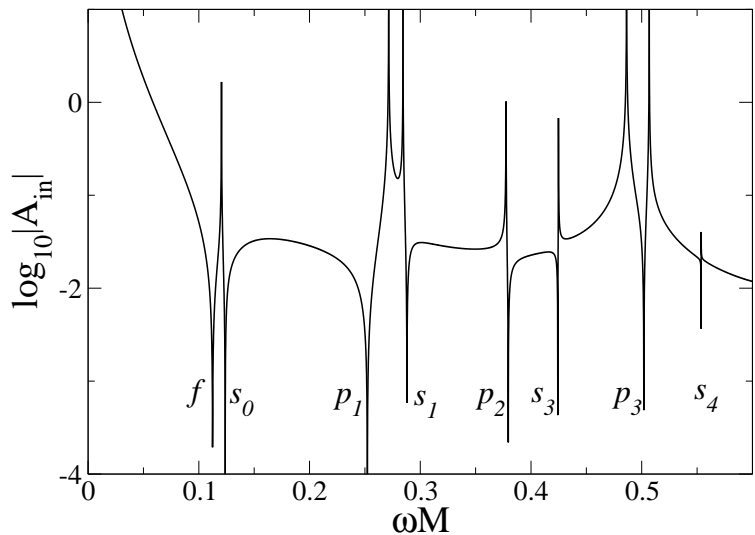}
\caption{This figure shows the asymptotic amplitude $A_{\rm in}$ as
a function of the (real) frequency $\omega M$ for our model~I. The slowly-damped 
QNMs of the star show up as zeros of $A_{\rm in}$, i.e. deep minima
in the figure. The first few ``ordinary'' and ``superfluid'' modes
are identified in the figure, cf. Table~III.}
\label{Ainplot}
\end{figure}

Since the w-modes are largely unaffected by superfluidity, and we do not 
expect pulsating g-modes, we focus our discussion on the ordinary 
fluid f- and p-modes as well as the superfluid s-modes.  We mentioned in 
the Introduction that there are two characteristics that distinguish 
the s-modes from their ordinary fluid counterparts: (i) counter-motion of 
the neutrons with respect to the protons, and (ii) a strong dependence on 
the parameters of entrainment.  We will leave until the next subsection 
the discussion on the effects of entrainment.  
Hence, we first consider the spectrum of fluid modes for a model
with ${\cal A} =0$. 
In Fig.~\ref{Ainplot} we 
provide 
a graph of the incoming wave amplitude (see Appendix III) versus the real 
part of the mode frequency for  our superfluid 
model~I.  As discussed in detail in Appendix III, a QNM 
corresponds to those particular solutions where there is no incoming wave 
at infinity.  Hence, the QNM frequencies 
correspond to the deep minima that can be seen in Fig.~\ref{Ainplot}.  The 
main feature to notice is the presence of twice as many slowly damped fluid 
modes as in the ordinary fluid case, cf. Fig.~6 of \cite{CLL99}.

\begin{table}
\caption{The oscillation frequency and the associated damping rate
(in terms of the real and imaginary parts of $\omega M$) for the first few
fluid pulsation modes of model~I. The w-modes for this model can be found 
in Table~III of \cite{CLL99}.}
\begin{tabular}{|c|cccccc|}
\hline
      mode & $f$ & $s_0$ & $p_1$ & $s_1$ & $p_2$ & $s_2$  \\ 
\hline 
      Re~$\omega M$ & 0.137 & 0.157 & 0.306 & 0.354 & 0.585 & 0.688 \\ 
      Im~$\omega M$ & $7.2 \times 10^{-5}$ & $4.2 \times 10^{-7}$ & 
      $6.6 \times 10^{-6}$ & $5.2 \times 10^{-6}$ & $3.2 \times 10^{-7}$ 
      & $5.5 \times 10^{-8}$ \\
\hline
\end{tabular} \label{freqtab}
\end{table}

In the earlier work of Comer, Langlois, and Lin \cite{CLL99}, there was 
no attempt at determining the gravitational-wave damping times for either 
the ordinary fluid or the superfluid modes.  
The reason for this was the difficulty in determining complex QNM frequencies
with imaginary parts that are orders of magnitude smaller than the real 
parts. To deal with this problem, we have developed a new method, which is
outlined in Appendix~III.  The 
results we have obtained for model~I are given in Table~\ref{freqtab}.  As one 
can see, the ordinary fluid modes have damping rates that are consistent 
with what is known from calculations of ordinary fluid neutron stars 
\cite{LD,DL}.  It should also be noticed that most of the superfluid 
modes have gravitational-wave damping rates that are similar to the 
high order p-modes. This is to 
be contrasted with recent statements in Ref.~\cite{SW00} that superfluid modes 
will not radiate gravitationally. In fact, in Section~\ref{nonrad} below
we will provide a proof that all QNMs of a superfluid star {\em must} 
radiate. This means that the superfluid modes could, at least in principle, 
be relevant for gravitational-wave
asteroseismology. This possibility will be discussed further in Section~VI.

So far we have mainly discussed the results obtained for model~I, for which 
there is no  envelope. When we turn to model~II, which has an ordinary 
fluid envelope of roughly one km, we find that the results do not 
change qualitatively. In particular, we do {\em not} find that new modes
arise because of the presence of the envelope. At first this may seem a 
little bit surprising. Especially since it is well known that an elastic crust 
supports several additional sets of modes. However, in our case 
the absence of new modes is due to the nature of the phase
transition at the core/envelope interface. We have chosen the 
phase-transition to be second order, which means that the number 
density of the protons remains smooth as the superfluid 
neutron component vanishes (at $r\to R_c$). Should we have 
taken the phase-transition to be first order, i.e. 
allowed for a  jump in the proton number density at $R_c$, 
our calculations would have unveiled a set of interface g-modes.

\subsection{The effect of entrainment --- avoided mode crossings}

The two main goals of the work presented in
this paper were: i) to allow for the presence
of a core/envelope transition, and thus in principle be able to consider
cases where the superfluid constituent is confined only to a part of the 
star, and ii) to determine how 
entrainment affects the QNM frequencies. As we will now discuss, 
the effect of entrainment can be considerable.  

We have carried out a series of calculations for model~II and 
entrainment parameters in the ``physical range'' $0.04<\epsilon < 0.2$. 
A sample of results are given in Table~\ref{freqtab2}.  
Listed there are the oscillation frequencies and associated damping rates 
for the first few pulsation modes of model~II and three different values 
of $\epsilon$.  First of all it is relevant to compare the 
results for the case of vanishing entrainment to those for model~I, cf. 
Table~\ref{freqtab}. Recall that the main difference 
between models~I and II is that the latter includes an ordinary
fluid envelope. Nevertheless, it is clear from the numerical 
results that the QNMs of the two models are qualitatively quite similar. 
When we turn to the effect of varying the entrainment parameter we find that 
the superfluid mode frequencies shift considerable, while the 
ordinary fluid modes remain virtually unchanged.  This is not a surprising
result given the discussion in Ref.~\cite{AC01a} and
 Eq.~(\ref{sfreq}). It is also relevant  to note that the 
gravitational-wave damping rates can be strongly 
affected by a change in $\epsilon$.

\begin{table}
\caption{The oscillation frequency and the associated damping rate
(in terms of the real and imaginary parts of $\omega M$) for the first few
fluid pulsation modes of model~II.  We show results for three different 
values of the entrainment parameter $\epsilon$. These correspond to the 
case of no entrainment as well as the upper and lower limits for the 
range that we take as ``physically realistic''. From this data we see that, 
while the ordinary fluid modes are hardly at all affected by the entrainment, 
the superfluid mode frequencies vary by as much as 10\% as $\epsilon$
is varied within the realistic range.}
\begin{tabular}{|c|c|cccccccc|}
\hline
      & mode & $f$ & $s_0$ & $p_1$ & $s_1$ & $p_2$ & $s_2$  & $p_3$ & 
      $s_3$\\ 
\hline 
      $\epsilon = 0$ & Re~$\omega M$ & 0.112 & 0.124 & 0.252 & 0.288 & 
      0.379 & 0.424 &0.502  &0.554 \\ 
      & Im~$\omega M$ & $4.9 \times 10^{-5}$ & $8.9 \times 10^{-6}$ 
      & $1.3 \times 10^{-5}$ & $1.4 \times 10^{-5}$ & $8.0 \times 10^{-8}$ 
      & $1.3 \times 10^{-7}$ & $1.5 \times 10^{-8}$ & $1.7 \times 10^{-9}$ 
      \\
\hline
      $\epsilon = 0.04$ & Re~$\omega M$ & 0.113 & 0.130 & 0.253  & 0.299 & 
      0.382 & 0.442 & 0.507 & 0.577 \\ 
      & Im~$\omega M$ & $5.3 \times 10^{-5}$ & $5.3 \times 10^{-6}$ &  
      $1.4 \times 10^{-5}$& $8.5 \times 10^{-7}$ & $9.2 \times 10^{-8}$ &  
      $1.2 \times 10^{-7}$ & $1.2 \times 10^{-8}$ & $4.3 \times 10^{-9}$ \\
\hline$
      \epsilon = 0.2$ & Re~$\omega M$ & 0.113 & 0.149 & 0.257  & 0.328 & 
      0.394 & 0.491 & 0.523 & 0.639 \\ 
      & Im~$\omega M$ & $5.5 \times 10^{-5}$ & $3.2 \times 10^{-6}$ &  
      $1.4 \times 10^{-5}$& $3.2 \times 10^{-7}$ & $1.0 \times 10^{-7}$ &  
      $1.2 \times 10^{-7}$ & $2.3 \times 10^{-9}$ & $1.2 \times 10^{-8}$ \\
\hline
\end{tabular} \label{freqtab2}
\end{table}

\begin{figure}[h]
\centering
\includegraphics[height=8cm,clip]{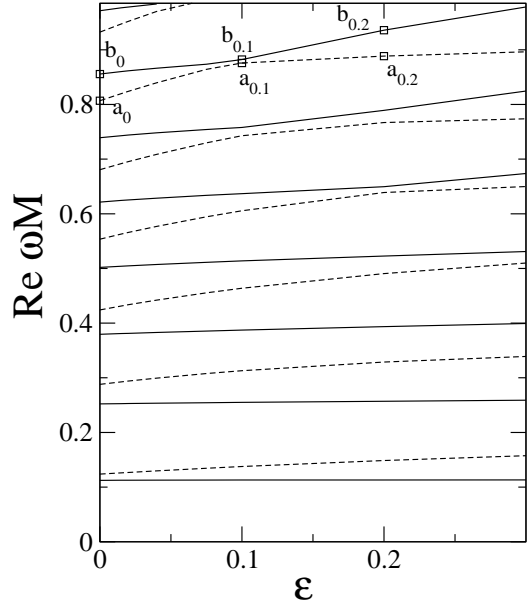}
\caption{
This figure shows how the frequencies of the fluid pulsation modes
for our model~II
vary with the entrainment parameter $\epsilon$. The modes shown 
as solid lines are such that the two fluids are essentially comoving 
in the $\epsilon \to 0$ limit, while the modes shown as dashed 
lines are countermoving. As is apparent from the data, the higher
order modes exhibit avoided crossings as $\epsilon$ varies. 
Recall that the range often taken as ``physically relevant'' is
$0.04 \le \epsilon \le 0.2$. We  indicate by $a_\epsilon$ and 
$b_\epsilon$ the particular modes for which the eigenfunctions are 
shown in Fig.~\ref{cross}.    }
\label{entrain}
\end{figure}

\begin{figure}[h]
\centering
\includegraphics[height=10cm,clip]{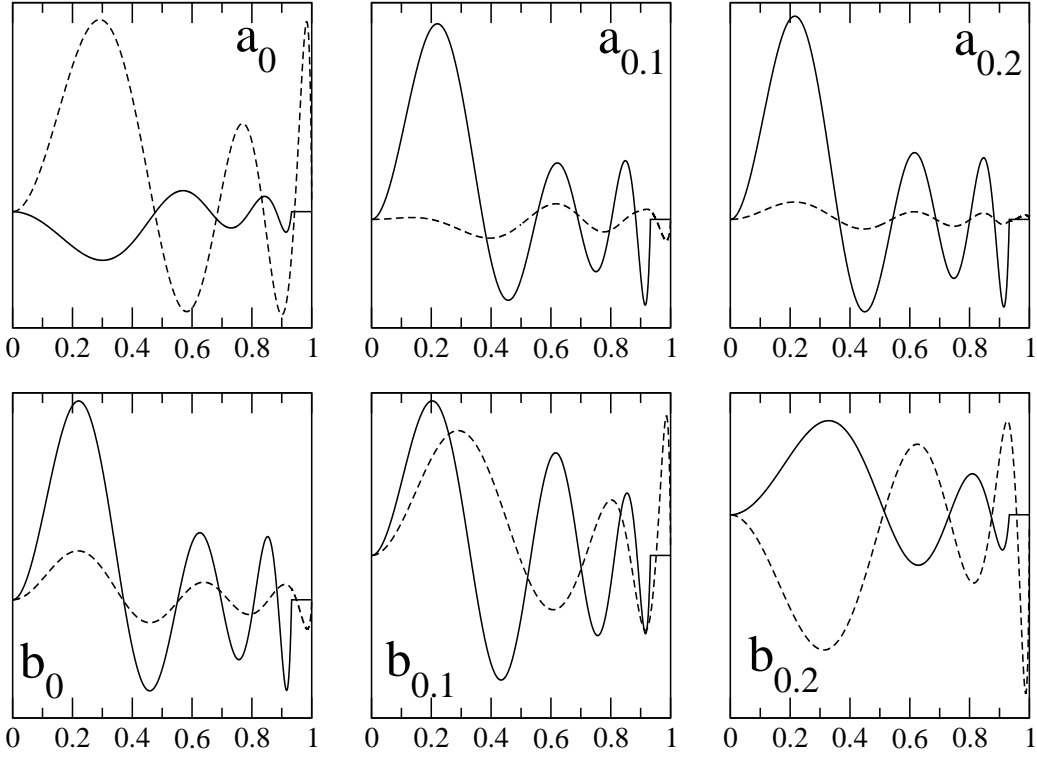} 
\caption{An illustration of the fact that the modes exchange properties
during an avoided crossing.  We  consider two modes, labelled by 
$a_\epsilon$ and $b_\epsilon$ (cf. Figure~\ref{entrain}). 
The mode eigenfunctions are represented by the two Lagrangian
number density variations, $\Delta n$ and $\Delta p$ (solid and dashed 
lines, respectively). 
 For mode $a$ 
the two fluids are essentially countermoving in the $\epsilon\to 0$ 
limit (it is a superfluid mode), while the two fluids comove for mode $b$ 
(it is similar to a standard p-mode). After the avoided crossing (which takes place
roughly 
at $\epsilon=0.1$) the two modes have exchanged properties.}
\label{cross}
\end{figure}

As we will now discuss, the
effect that a varying entrainment has on the damping rate of the modes
can be understood from the results illustrated in 
Figs.~\ref{entrain} and \ref{cross}.  
Fig.~\ref{entrain} shows how the 
oscillation frequencies for the first few modes change as the entrainment 
parameter is varied within the acceptable range.
 The modes that have ordinary 
fluid behavior, i.e.~for which 
the neutrons and protons ``flow together'', in the limit of vanishing
entrainment are shown as solid lines in the figure whereas the 
superfluid modes, where the 
neutrons and protons are largely counter-moving, are given by the dashed 
lines.  The main feature of the figure is the presence of so-called 
avoided crossings. For the higher order modes 
(near the top of the figure) there are points 
in the $(\epsilon,{\rm Re}~\omega M)$ plane where the solid and dashed 
lines approach each other, but rather than crossing they diverge 
from each other.  An interesting aspect of 
these avoided crossings can be gleaned from Fig.~\ref{cross}, which 
shows the Lagrangian variations $\Delta n$ and $\Delta p$ for 
the particular modes that correspond to the points $a_\epsilon$ and 
$b_\epsilon$, $\epsilon = 0, 0.1,0.2$, of Fig.~\ref{entrain}.  For the 
mode $a$ the two fluids are largely counter-moving as $\epsilon \to 0$, 
and for mode $b$ the two fluids are then comoving.  However, as $\epsilon \to 
0.2$ we see that the modes have changed character in that it is now mode $a$ 
that is comoving and mode $b$ that is counter-moving.  This exchange of 
character of modes is a  characteristic of avoided crossings 
familiar from other problems in stellar pulsation theory \cite{japan}.
The presence of avoided crossings between stellar pulsation modes is 
familiar from many other situations, but we believe that our results provide
the first  hard evidence for the presence of this phenomenon in 
superfluid stars.

From general post-Newtonian arguments, one would expect the co-moving modes
to radiate gravitational waves more efficiently than the counter-moving
superfluid modes. Thus it is not surprising to find that the damping 
rate of a superfluid mode increases as entrainment is varied and 
an avoided crossing is approached. In the present context, this effect is 
probably not distinct enough to be of great relevance but one can argue 
that it may be of great importance in closely related problems.   
As has been argued recently \cite{AC01a}, avoided crossings 
may be at the heart of recent calculations of 
the effect of superfluidity on the r-mode instability.  Lindblom and 
Mendell \cite{LM00} have analyzed the effect of mutual friction 
damping on the r-modes using the same entrainment model as we 
employ here.  They 
found that mutual friction was, in general, not effective at damping 
the r-modes. However, they also found 
(cf.~their Fig.~6) that the mutual friction damping 
time could be very small for particular values of $\epsilon$.  We 
believe that a proper explanation of this result can be obtained via the 
avoided crossings phenomenon.  The basic idea is that mutual friction 
should be most effective whenever the neutrons and protons are 
counter-moving, as with the superfluid modes.  Andersson and Comer 
\cite{AC01a} have shown that there are two sets of r-modes, that are quite 
analogous to the ordinary fluid and superfluid modes discussed in this paper 
in the sense that one set has the neutrons and protons comoving whereas the 
other set has them countermoving.  Although avoided crossings between these 
two classes of r-modes in a superfluid star have not yet
been demonstrated, we believe that our current results provide 
strong support for the idea by demonstrating the presence of 
avoided crossings in a very closely connected situation. 
 We thus assert that the particular values of $\epsilon$ for with 
Lindblom and Mendell find  
small mutual friction damping times correspond to stellar models
for which the two classes of superfluid r-modes are 
close to an avoided crossing. 
The veracity of this argument remains to be confirmed by detailed calculations
that we plan to carry out in the near future.

\section{Are there non-radiative modes in superfluid neutron stars?} 
\label{nonrad}

In this Section we digress somewhat  
and focus our attention on a question of principle:
Is it possible to have QNMs that do not 
radiate gravitationally in a superfluid star?
The question is motivated by the simple fact that, while every 
non-radial motion 
induced in a one-fluid system must lead to the 
emission of gravitational waves,  the situation could conceivably 
be different in the two-fluid case. One can imagine the 
possibility that the two fluids move in such a way that the 
average mass-density flux vanishes identically. In fact, as we have 
discussed elsewhere \cite{AC01a}, the superfluid modes 
are essentially of this nature and it has  been argued by 
Sedrakian and Wasserman \cite{SW00} that this means that these modes will not radiate. 
Of course, one must not forget that, in the post-Newtonian picture, 
gravitational waves are associated with both mass- and current multipoles.
Thus even the extreme case of a two-fluid oscillation 
such that the mass-multipoles vanish identically
is likely to radiate through the induced currents. Thus it may be very 
difficult to set the two fluids into motion without generating
gravitational waves.  
And this is, indeed, the way that it turns out. As we will prove 
below, a two-fluid star can (in general) have no non-radiative modes. 

We consider the superfluid perturbation equation in the special 
case where the metric is left completely unperturbed, i.e. when 
no gravitational radiation is created. 
Assuming a given, fixed non-rotating background,  we can obtain the
corresponding 
field equations by setting to zero the three metric 
perturbation $H_0$, $H_1$, and $K$. This results in the following nine 
equations for six matter variables, cf. Eqns (\ref{cnstrt})-(\ref{xpprime}),

\begin{eqnarray}
  0 &=& \Xn + \Xs \ ,  \label{cnstrt2}\\
  && \cr
  0 &=& \mu \n \Vn + \chi \p \Vs \ , \label{h1prime2} \\
  && \cr
  0 &=& \mu \n \Wn + \chi \p \Ws \ , \label{Kprime2} \\
  && \cr
  \Wn^{\prime} &=& - e^{\lambda/2} {l (l + 1)\over r} \Vn - \left({l + 1 
    \over r} + {n^{\prime} \over \n}\right) \Wn \cr
  &&
    + {\c00 \over \n^2 \D00} \left[e^{(\lambda - \nu)/2} r \Xn + 
    n^{\prime} \left(\b00 \n \Wn + \a00 \p \Ws\right)\right] - \cr
  &&
    {\a00 \over \n \p \D00} \left[e^{(\lambda - \nu)/2}r \Xs + \p^{\prime} 
    \left(\a00 \n \Wn + \c00 \p \Ws\right)\right] \ , \label{wnprime2} \\
  && \cr
  \Ws^{\prime} &=& - e^{\lambda/2} {l (l + 1) \over r} \Vs - \left({l + 1 
    \over r} + {\p^{\prime} \over \p}\right) \Ws \cr
  &&
    + {\b00 \over \p^2 \D00} \left[e^{(\lambda - \nu)/2} r \Xs + 
    \p^{\prime} \left(\c00 \p \Ws + \a00 \n \Wn\right)\right] - \cr
  &&
    {\a00\over \n \p \D00} \left[e^{(\lambda - \nu)/2}r \Xn + \n^{\prime} 
    \left(\a00 \p \Ws + \b00 \n \Wn\right)\right] \ , \label{wpprime2} \\
  && \cr
  \Xn^{\prime} &=& - {l \over r} \Xn + {l (l + 1) \over r^2} e^{\nu/2} 
    \n^{\prime} \left(\b00 \n \Vn + \a00 \p \Vs\right) \cr
  &&
    - e^{(\lambda - \nu)/2} {\omega^2 \over r} \n \left(\B \n \Wn + \A \p 
    \Ws\right) \cr
  &&
    + e^{-(\lambda - \nu)/2} \left[- {\n^{\prime} \over r} \left(
    \b00^{\prime} \n \Wn + \a00^{\prime} \p \Ws\right) + \right. \cr
  &&
    \left.\left({2 \n^{\prime} \over r^2} + {\lambda^{\prime} - 
    \nu^{\prime} \over 2 r} \n^{\prime} - {\n^{\prime \prime} \over r}
    \right) \left(\b00 \n \Wn + \a00 \p \Ws\right)\right] \ , 
    \label{xnprime2} \\
  && \cr
    \Xs^{\prime} &=& - {l \over r} \Xs  + {l (l 
    + 1) \over r^2} e^{\nu/2} \p^{\prime} \left(\c00 \p \Vs + \a00 \n 
    \Vn\right) \cr
  &&
    - e^{(\lambda - \nu)/2} {\omega^2 \over r} \p \left(\C \p \Ws + \A \n 
    \Wn\right) \cr
  &&
    + e^{- (\lambda - \nu)/2} \left[- {\p^{\prime} \over r} \left(
    \c00^{\prime} \p \Ws + \a00^{\prime} \n \Wn\right) + \right. \cr
  &&
    \left.\left({2 \p^{\prime} \over r^2} + {\lambda^{\prime} - 
    \nu^{\prime} \over 2 r} \p^{\prime} - {\p^{\prime \prime} \over r}
    \right) \left(\c00 \p \Ws + \a00 \n \Wn\right)\right] 
    \ , \label{xpprime2}
\end{eqnarray}
with 
\beq
    \Xn \equiv \n e^{-\nu/2} \omega^2 \left(\B \n \Vn + \A \p \Vs\right)
     - e^{(\nu - \lambda)/2} {\n^{\prime} \over r} \left(\b00 \n \Wn + 
    \a00 \p \Ws\right) \ , \label{xn2}
\eeq
and
\beq
    \Xs \equiv \p e^{-\nu/2} \omega^2 \left(\C \p \Vs + \A \n \Vn\right)
    - e^{(\nu - \lambda)/2} {\p^{\prime} \over r} \left(\c00 \p \Ws + 
    \a00 \n \Wn\right) \ . \label{xs2}
\eeq
It is clear that, unless some of these 
equations are  linearly dependent the problem is overdetermined and we cannot 
have a non-trivial solution. 
In order to show that the problem is in general overdetermined,  we use 
the first three equations above to 
rewrite $W_\p$, $V_\p$, and $X_\p$ in terms of the neutron variables 
$W_\n$, $V_\n$, and $X_\n$, and then substitute these into the remaining 
six equations.  Remarkably, 
one finds that Eqs.~(\ref{xnprime2}) and (\ref{xpprime2}) yield exactly 
the same result when this substitution is done. One also finds that 
Eqs.~(\ref{xn}) and (\ref{xs})  yield the same result after the 
substitution.  We are thus left with Eqs.~(\ref{wnprime2}) and 
(\ref{wpprime2}). After the substitution, 
these equations yield the following two
equations for $W_\n$:
\begin{eqnarray}
   W^{\prime}_\n &=& - e^{\lambda/2} {l (l + 1) \over r} V_\n - {l + 1 
      \over r} W_\n + {\a00 \over \D00} \left(\a00 - {\mu \over \chi} 
      \c00\right) \left({\n^{\prime} \over \n} - {\p^{\prime} \over 
      \p}\right) W_\n \cr
      &&+ e^{(\lambda - \nu)/2} r \left({\c00 \p + \a00 \n 
      \over \p \n^2 \D00}\right) X_\n \ , \label{weq1} \\
    && \cr
   W^{\prime}_\n &=& - e^{\lambda/2} {l (l + 1) \over r} V_\n - {l + 1 
      \over r} W_\n - {\b00 \over \D00} \left(\c00 - {\chi \over \mu} 
      \a00\right) \left({\n^{\prime} \over \n} - {\p^{\prime} \over 
      \p}\right) W_\n \cr
      &&+ e^{(\lambda - \nu)/2} r {\chi \over \mu} \left({\b00 \n + \a00 
      \p \over \p \n^2 \D00}\right) X_\n \label{weq2} \ .
\end{eqnarray}
In the manipulations we have used repeatedly the background equations (\ref{bgndfl_c})
and
\beq
    \mu^{\prime} = - {1 \over 2} \mu \nu^{\prime} \quad , \quad 
    \chi^{\prime} = - {1 \over 2} \chi \nu^{\prime} \ .
\eeq

In order to complete the argument in a clear way, we 
restrict ourselves to the case of vanishing 
entrainment and separable equations of state, i.e. we concentrate on the 
case $\A=\a00=0$. 
This is not at all necessary, but it simplifies the analysis considerably
since the involved relations are much simpler. 
Taking the difference between (\ref{weq1}) and (\ref{weq2})
we require 
\beq
\B_0^0 \C_0^0 \left( { n^\prime \over n} - {p^\prime \over p}\right) W_n 
+ { e^{(\lambda - \nu)/2} r \over p n^2} \left( \C_0^0 p - { \chi \over \mu} \B_0^0 n  \right) X_n = 0
\eeq
in order for the problem not to be over-determined. Using the definition of 
$X_n$ we can rewrite this equation as 
\beq
e^{\lambda/2 - \nu} r \omega^2 \B \left( \C_0^0 - { n \chi \over p \mu} 
\B_0^0  \right) V_n + { \B_0^0 \over p} \left( n^\prime \B_0^0 { \chi \over \mu} - p^\prime \C_0^0\right) W_n  = 0 \ .
\eeq
Now using the fact that $\chi \equiv \C p$ and $\mu \equiv \B n$ in the case
of vanishing entrainment, we have
\beq
e^{\lambda/2 - \nu} r \omega^2  \left(\B \C_0^0 -
\C \B_0^0  \right) V_n + { \B_0^0 \over \B } \left( \C \B_0^0 { n^\prime 
\over n} - \B \C_0^0 { p^\prime \over p } \right) W_n =
e^{\lambda/2 - \nu} r \omega^2  \left(\B \C_0^0 - \C \B_0^0  \right) V_n = 0 \ .
\eeq
To arrive at the last equality we have employed the background identities
(\ref{bgndfl_c}) again. 
Clearly, we are now left with two possibilities. Either $V_n$ vanishes, or
the equation of state must be such that
\beq
\B \C_0^0 - \C \B_0^0 = 0 \ .
\eeq
In the first case one can show that the corresponding solution for $W_n$ is
\beq
W_n \propto{ 1 \over n r^{l+1}} \ .
\eeq
 This solution is clearly physically unacceptable 
since it diverges at the centre of the star. In other words, if $V_n=0$ we 
must also have $W_n=0$, i.e. the trivial solution. 
It is easy to show that the second case requires that the master function 
be such that
\beq
8p^2{ \partial \Lambda \over \partial n^2} {\partial \over \partial p^2} \left(
{  \partial \Lambda \over \partial p^2} \right) -
8n^2{ \partial \Lambda \over \partial p^2} {\partial \over \partial n^2} \left(
{  \partial \Lambda \over \partial n^2} \right) = 0 \ .
\eeq
This is clearly a very particular form. We have thus proven that the QNMs 
of a superfluid star {\em must} radiate unless the equation of state
belongs to a very special class. The obvious exceptions are i) when the 
master function takes the form
\beq
\Lambda = \sigma_n n^2 + \sigma_p p^2
\eeq
and ii) whenever $\Lambda$ depends on $n$ and $p$ in identical ways.

It is worth pointing out that the calculation in this section was carried 
out within the Regge-Wheeler gauge, and given possible gauge issues
one may worry 
that this means that the result is of limited validity. 
However, the result will hold in general since we can easily construct
gauge-invariant quantities (such as the Zerilli function) 
that represent the gravitational-wave degrees of freedom
from the metric perturbations calculated in any particular gauge.  
In our case it is trivial to see that, if all 
metric perturbations vanish identically the Zerilli function will be identically
zero and no gravitational-waves will emerge from the system.

\section{Detectable Gravitational Wave Signals?} \label{gws}

Given that a new generation of gravitational-wave 
detectors are likely to be operating at their projected 
levels sensitivity within  the next few months,  
it is appropriate to conclude this paper with a brief discussion of a 
possible future application of our results. Suppose that
the various modes of a superfluid neutron star were excited
to an amplitude such that the associated gravitational-wave signal could be 
detected. To what extent would it then be possible to solve the inverse
problem and deduce information concerning the superfluid parameters?
In other words, can we hope to use ``gravitational-wave asteroseismology''
to probe the superfluid interior of mature neutron stars?  
This question has recently been discussed by two of us \cite{AC01b}, and 
therefore we only provide a brief background here. 

As has already 
been discussed by Kokkotas et al \cite{astero2}, the detection of
gravitational-wave signals from pulsations in newly born neutron stars
could be used to infer the mass and 
radius of the star. This 
information would put strong constraints on the supranuclear 
equation of state. However, this argument relies on releasing 
an energy equivalent to something like $10^{-5} M_\odot c^2$ through the 
QNMs. This may not be unreasonable for the wildly pulsating 
object formed through a strongly asymmetric supernova collapse, but it 
is difficult to think of a mechanism whereby the oscillations of a
mature (and thus superfluid) neutron star core will be excited to 
a similar level. Instead, we take as a ``reasonable'' order of magnitude estimate
the energy associated with a typical pulsar glitch.   
The released energy can then be estimated as
\begin{equation}
\Delta E \approx I \Omega \Delta \Omega \approx (10^{-6}-10^{-8}) I 
\Omega^2
\end{equation}
where $\Omega = 2\pi /P$ is the rotation rate of the star, and $P$ is the
observed pulsar period.
In this formula it is appropriate to use the moment of inertia 
$I\sim 10^{45}$~gcm$^2$ 
of the entire star, since the 
spin-up incurred during the glitch remains on timescales that are 
much longer than the estimated coupling timescale between the crust 
and the core fluid. By combining the above formula with the 
data for typical glitches in the Crab and Vela pulsars, cf. 
Table~\ref{glitchtab}, we arrive at estimates of  
 the energy associated with typical 
glitches that accord well with suggestions in the literature
 \cite{franco00,hart00}. 

\begin{table}[h]
\caption{Data for archetypal glitching pulsars.
\label{glitchtab}}
\begin{tabular}{|l|ccc|r|}
\hline
PSR & $P$ (ms) & $d$ (kpc) & $\Delta \Omega / \Omega$ & $\Delta E/M_\odot 
c^2$ \\
\hline
Crab & 33 & 2 & $10^{-8}$ & $2\times10^{-13}$ \\
Vela & 89 & 0.5 & $10^{-6}$ & $3\times10^{-12}$\\
\hline
\end{tabular}
\end{table}
 
As is evident from  
Table~\ref{glitchtab}, we expect a glitch to be associated with energies of the 
order of $10^{-13}-10^{-12}M_\odot c^2$. In the following we will assume 
that a similar energy is channeled through the various 
pulsation modes. This then allows us to use the formulas obtained by 
Kokkotas et al \cite{astero2} to estimate, for a given detector configuration, the 
attainable signal-to-noise ratio.

Assume that the gravitational-wave signal from a
neutron star pulsation mode takes the
form of a damped sinusoidal, i.e.
\begin{equation}
h(t) = {\cal A} e^{-(t-T)/t_d} \sin [ 2\pi f (t-T)] \quad \mbox{ for } 
       t > T
\end{equation}
where $f$ is the frequency of the QNM, $t_d$ is its characteristic damping time, 
and  $T$ is the arrival time of the signal at the detector (thus $h(t)=0$ for $t<T$). 
Using standard results for the gravitational-wave flux \cite{astero2},
the amplitude ${\cal A}$ of the signal can be expressed in terms of the 
total energy radiated through the mode:
\begin{equation}
{\cal A} \approx 7.6\times 10^{-24} \sqrt{{\Delta E_\odot \over 
10^{-12} } {1 \mbox{ s} \over t_d}}  
 \left( {1 \mbox{ kpc} \over d } \right) 
\left({ 1 \mbox{ kHz} \over f} \right)  \ .
\end{equation}
where $\Delta E_\odot = \Delta E/M_\odot c^2$.
The signal-to-noise ratio for this signal can be 
estimated from \cite{astero2}
\begin{equation}
\left({S \over N} \right)^2 = { 4Q^2 \over 1+4Q^2} {{ \cal A}^2 t_d 
\over 2S_n}
\label{sign}\end{equation}
where the ``quality factor'' is $Q=\pi f t_d$ and 
$S_n$ is the spectral noise density of the detector. 

\begin{figure}[tbh]
\centering
\includegraphics[height=6cm,clip]{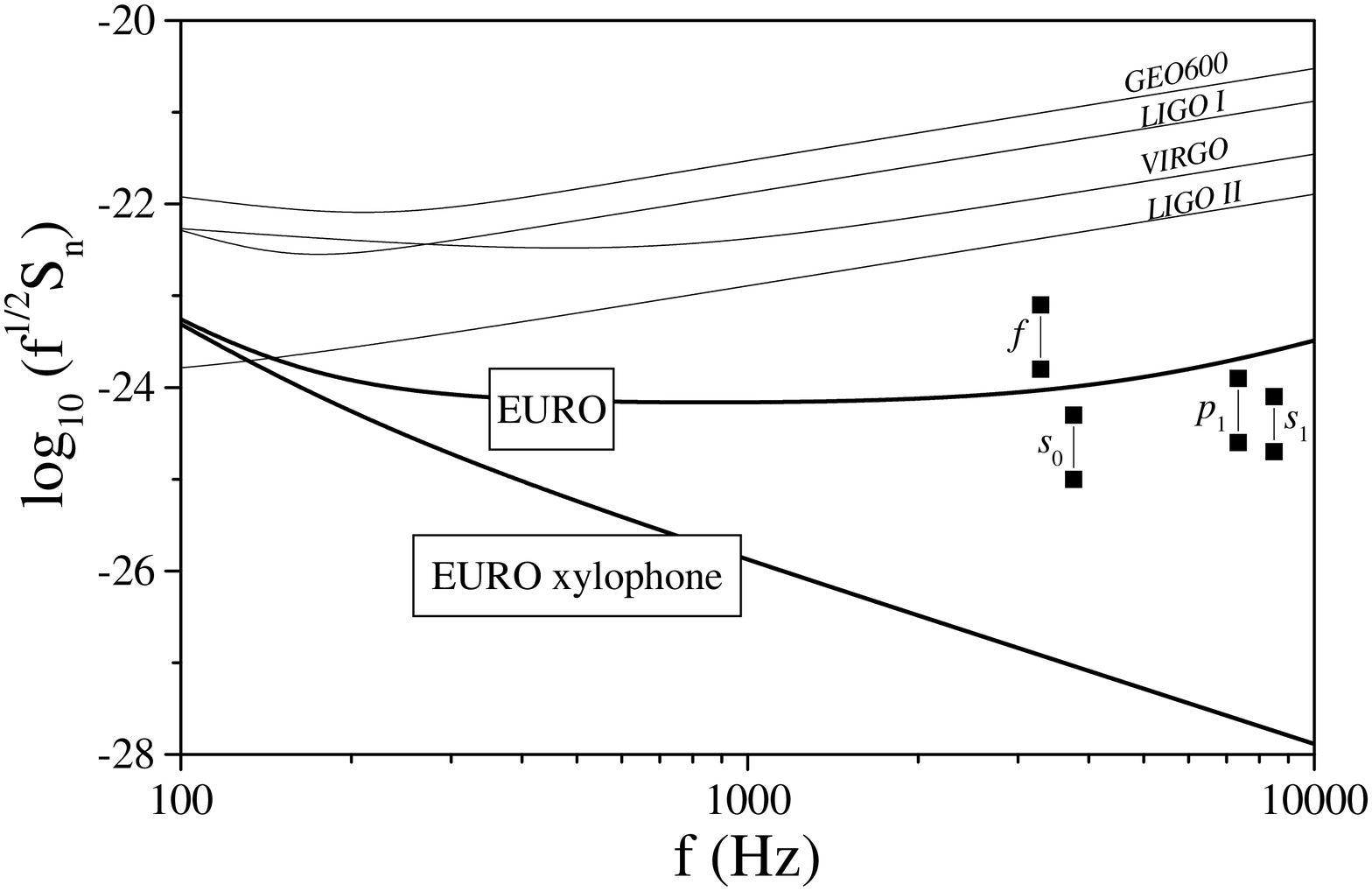}
\caption{The spectral noise density for the new generation of 
laser-interferometric gravitational-wave detectors that will come 
online in the next few years (thin lines) is compared to 
speculative estimates for the futuristic EURO detector (solid lines).  A key 
feature of this advanced configuration is that it may operate several 
narrowbanded interferometers as a xylophone, thus reaching high 
sensitivity at kHz frequencies. We also indicate the effective 
gravitational-wave amplitudes from the glitch-induced mode-oscillations
discussed in the text.}
\label{figure1}\end{figure}

We can now combine these estimates with the QNMs from (say) 
Table~\ref{freqtab}. The relevant frequencies and damping times
in kHz and ms, respectively, are given in Table~\ref{ftab}.
When we compare the obtained estimates to the new generation 
of large-scale interferometric detectors it immediately becomes 
clear that these signals would be too weak to be detected. 
This is illustrated in 
Figure~\ref{figure1} where  
we compare the dimensionless strain  $\sqrt{f} S_n$ for various 
detector configurations to the estimated strain caused by the 
QNM signals. 

\begin{table}
\caption{The frequency and damping rate for the first few modes of 
our Model~I (which is identical to model~2 of Comer et al \cite{CLL99}).
We also show the gravitational-wave
signal-to-noise ratios resulting from 
the ``glitch model'' discussed in the main text. The results correspond to 
an advanced EURO detector with (case 1) and without (case 2) photon
shotnoise, respectively. The 
lower estimate is for a Crab glitch while the upper estimate follows 
from the Vela data. 
\label{ftab}}
\begin{tabular}{|l|cc|cc|}
\hline
Mode & $f$ (kHz) & $t_d$ (s) & Case 1 & Case 2\\
\hline
$f$ & 3.29 & 0.092  & 0.4---6 & 300---$4.7\times10^3$\\
$p_1$ & 7.34 & 1.01 & 0.08---1.2 & 680---$1\times10^4$ \\
\hline
$s_0$ & 3.76 & 15.75 & 0.3---4.8 & 350---$5.4\times10^3$ \\
$s_1$ & 8.49 & 1.29 & 0.06---0.9 & 780---$1.2\times 10^4$\\
\hline
\end{tabular}
\end{table}

This does not, however, mean that we should give 
up on the main idea behind this analysis. We simply have to 
acknowledge that we are likely to require significant 
improvements in technology if we are to be able to make this
into a viable approach. But it seems inevitable that 
the available technology will improve over the next decades.  
 In fact,  various groups are already 
discussing possible  improvements in detector sensitivity
that may be achievable in the future. In order to 
illustrate the levels that are being discussed, we consider the
so-called EURO detector, for which the sensitivity has been 
estimated by Sathyaprakash and Schutz 
(for further details see \cite{euro}).  We 
consider two possible configurations: In the first, the sensitivity at
high frequencies is limited by the photon shot-noise, while 
the second configuration reaches beyond this limit by running 
several narrowbanded (cryogenic) interferometers as a ``xylophone''.
The corresponding noise-curves are illustrated, and 
compared to the current generation of interferometers, in 
Figure~\ref{figure1}.  
It is immediately clear from 
Figure~\ref{figure1} that a 
EURO detector would be a superb
instrument for studying pulsating neutron stars. 
This means that 
previously suggested strategies \cite{astero2}  
for unveiling the supranuclear equation of 
state may eventually be put to the test. In fact, as is clear from 
the estimates in Table~\ref{ftab}, where we list the  
signal-to-noise ratio estimated from (\ref{sign}),  
 one should also be able to 
detect the superfluid oscillation modes. 
The various modes would be marginally detectable given this level 
of excitation and a third generation
detector limited by the photon shotnoise. 
If this limit can be surpassed by configuring several 
narrowbanded interferometers as a xylophone, the achievable 
signal-to-noise ratio will be excellent. 
Hence, it seems plausible that 
one could infer the parameters of neutron star superfluidity. As an obvious
``by product'' one might hope to  
shed light on the mechanism for pulsar glitches.

We can also confirm that, provided that the modes can be detected, 
the oscillation frequencies can be extracted with good accuracy
from the data. By combining  Eqns.~(11) and (12) from Kokkotas et al \cite{astero2}
with (\ref{sign}) one can obtain a formula for the signal-to-noise required 
to determine the mode frequency with a relative error $\sigma_f/f$. 
We get (since $Q>>1$ for the modes under consideration) a relation
\beq
\left( { S \over N} \right) \approx 0.33 \left( { 1 \mbox{ s} \over t_d} \right) 
\left( { \mbox{1 kHz} \over f}\right) \left({ \sigma_f \over f} \right)^{-1}
\eeq
From this relation we can deduce that a detection with signal-to-noise
of (say) 10 would enable one to infer the fluid mode frequencies
with an accuracy of the order of a percent or so. With this level
of precision one should be able to  distinguish clearly between the 
``normal fluid'' f and p-modes and the superfluid s-modes.  In other 
words, we would have the information required not only to infer the 
mass and radius of the star \cite{astero2}, we could also hope to 
constrain the
parameters of neutron star superfluidity.  

\section{Conclusions} \label{con}

The main motivation behind the work presented in this paper
is the fact that neutron star physics is not adequately modelled
within Newtonian gravity. It is well known that Newtonian results differ 
greatly from the correct relativistic ones already at the 
level of determining the mass and radius of a star with a 
prescribed central density from a given ``realistic'' equation of state.
As far as the QNMs of the star are concerned, Newtonian studies are
extremely useful since they help us understand the physics of
different classes of modes. But at the same time 
it is clear that if we require a detailed
model of the oscillation spectrum of an astrophysical neutron star
we must approach the problem from the relativistic point of view.
This is particularly crucial if we are interested in the 
gravitational-wave damping rates. 
 Furthermore, since mature neutron stars are likely 
to contain several superfluid 
components, it is important that we develop a framework for modelling 
multi-fluid systems in full general relativity. 
The present paper represents significant progress towards this goal. 

We have considered a core-envelope model, with superfluid neutrons 
being present only in the core of the star while the outer region is 
composed of an ordinary fluid. Our model includes a simple, yet reasonable, 
model for entrainment and we have shown how neighbouring QNMs undergo 
avoided crossings as the entrainment parameter is varied. 
Finally, we have ruled out the possibility that a two-fluid star could have 
non-radiative pulsation modes, and discussed the possible detection 
of gravitational-wave signals from oscillating superfluid neutron stars. 

In the near future, our aim is to turn our attention to the oscillations
of rotating superfluid stars. This is an exciting problem area, since
various modes of oscillation may be driven unstable by the emission 
of gravitational waves. Of particular recent interest are the so-called
r-modes, and the damping due to superfluid mutual friction. 
In a recent work, Lindblom and Mendell \cite{LM00} showed that mutual 
friction was effective at damping unstable r-modes only for very particular values of 
the entrainment parameter.  Andersson and Comer \cite{AC01a} have speculated that this 
peculiar behavior might be due to the existence of avoided crossings between 
two classes of r-modes (analogous to the two classes of acoustic fluid 
modes discussed in the 
present paper). The results of the present paper  lend support 
to this idea. Yet it is clear that our understanding of the various 
involved issues is far from complete, and that a considerable amount 
of work remains to be done before we can model the dynamics of rotating 
superfluid neutron stars in a ``realistic'' way. 

\acknowledgments

We thank Brandon Carter and John Clark for their excellent answers to 
several questions that arose as this work was being pursued.  We are also
grateful to Reinhard Prix for useful discussions and a careful reading of
a draft version of this paper. NA 
gratefully acknowledges support from the Leverhulme Trust via a Prize 
Fellowship, PPARC grant PPA/G/1998/00606 as well as the EU Programme 
'Improving the Human Research Potential and the Socio-Economic Knowledge 
Base' (Research Training Network Contract HPRN-CT-2000-00137).  GLC 
gratefully acknowledges partial support from a Saint Louis University 
SLU2000 Faculty Research Leave award, and EPSRC visitors grant 
GR/R52169/01 in the UK.

\section*{Appendix~I: The Junction Conditions}

In this Appendix will be presented a geometrical approach to deriving 
the junction conditions that must be used to smoothly join together 
an inner superfluid core with an exterior normal fluid envelope.  The 
junction conditions will be obtained via an analysis of the first and 
second fundamental forms associated with the (in general, timelike) 
hypersurfaces given by the level surfaces of the generalized pressure 
$\Psi$ (cf.~Eq.~(\ref{press})) \footnote{One might consider using the 
level surfaces of other scalar quantities, such as either of the particle 
number densities, or the Master function $\Lambda$.  However, we believe 
that the pressure is the most natural choice from both mathematical and 
physical points of view.}.  If there are no ``delta-function-like'' 
discontinuities in the pressure, then the first and second fundamental 
forms will be continuous throughout the star \cite{MTW}.  Turning the 
problem around, by demanding the continuity of the first and second 
fundamental forms a smooth joining at the core/envelope interface 
can be achieved.  

Consider that the level surfaces of $\Psi$ are timelike, i.e.~the normal 
to these surfaces is given by 
\beq
    {\cal N}^{\mu} = {g^{\mu \nu} \nabla_{\nu} \Psi \over \sqrt{\nabla_{\mu} 
              \Psi \nabla^{\mu} \Psi}} \ ,
\eeq
where ${\cal N}_{\mu} {\cal N}^{\mu} = 1$.    The so-called 
first fundamental form $\gamma_{\mu \nu}$ (i.e.~the induced three-metric) 
of these level surfaces is
\beq
     \gamma_{\mu \nu} = \perp^{\sigma}_{\mu} \perp^{\tau}_{\nu} 
                        g_{\sigma \tau} \ ,
\eeq 
where the ``perp'' operator $\perp^{\sigma}_{\mu}$ is given by
\beq
     \perp^{\sigma}_{\mu} = \delta^{\sigma}_{\mu} - {\cal N}^{\sigma} {\cal N}_{\mu} 
     \ .
\eeq
The second fundamental form $K_{\mu \nu}$ (i.e.~the extrinsic curvature) 
of the level surfaces is defined as
\beq
     K_{\mu \nu} = - \perp^{\sigma}_{\mu} \perp^{\tau}_{\nu} 
                   \nabla_{(\sigma} {\cal N}_{\tau)} \ ,
\eeq
where the parentheses imply symmetrization of the indices.

Let us consider  the pressure to be of form
\beq
     \Psi(t,r,\theta) = \Psi_0(r) + \delta \Psi(t,r,\theta) \ .
\eeq
The components of the unit normal ${\cal N}^{\mu}$ are thus found to be
\begin{eqnarray}
       {\cal N}^0 &=& - e^{\lambda/2 - \nu} {\delta \dot{\Psi} \over 
               \Psi_0^{\prime}} + e^{- (\lambda/2 + \nu)} \delta g_{0 1} 
           \ , \cr
           && \cr
      {\cal N}^1 &=& e^{- \lambda/2} \left(1 - \delta g_{1 1}/2 e^{\lambda}
               \right) \ , \cr
           && \cr
       {\cal N}^2 &=& {e^{\lambda/2} \over r^2} {\delta \Psi_{, \theta} \over 
               \Psi_0^{\prime}} 
           \ , \cr
           && \cr
       {\cal N}^3 &=& 0 \ .
\end{eqnarray}
The non-zero components of the first fundamental form are
\begin{eqnarray}
     \gamma_{0 0} &=& - e^{\nu} + \delta g_{0 0} \ , \quad \gamma_{0 1} 
     = \delta g_{0 1} - e^{\lambda} {\delta \dot{\Psi} \over 
     \Psi_0^{\prime}} \ , \quad \gamma_{12} = - e^{\lambda} 
    {\delta \Psi_{, \theta} \over \Psi_0^{\prime}} \ , \cr
     && \cr
     \gamma_{2 2} &=& r^2 + \delta g_{2 2} , \quad \gamma_{3 3} = 
     {\rm sin}^2 \theta r^2 + \delta g_{3 3} \ ,
\end{eqnarray}
and we also find
\begin{eqnarray}
     K_{0 0} &=& {\nu^{\prime} \over 2} e^{\nu - \lambda/2} - 
     e^{\lambda/2} {\delta \ddot{\Psi} \over \Psi^{\prime}_0} + {1 
     \over e^{\lambda/2}} \delta \dot{g}_{0 1} - {1 \over 2 
     e^{\lambda/2}} \delta g^{\prime}_{0 0} + {\nu^{\prime} 
     \over 4} e^{\nu - 3 \lambda/2} \delta g_{1 1} \ , \cr
     && \cr
     K_{0 1} &=& {\nu^{\prime} \over 2} \left(e^{\lambda/2} {\delta 
     \dot{\Psi} \over \Psi_0^{\prime}} - e^{- \lambda/2} \delta 
     g_{0 1}\right) , \quad K_{0 2} = - e^{\lambda/2} \left({\delta 
     \dot{\Psi}_{,\theta} \over \Psi_0^{\prime}} - {1 
     \over 2 e^{\lambda}} \delta g_{0 1,\theta}\right) \ , \cr 
     && \cr
     K_{1 2} &=& {e^{\lambda/2} \over r} {\delta \Psi_{,\theta} \over 
     \Psi^{\prime}_0} , \quad K_{2 2} = - {r \over e^{\lambda/2}} - 
     e^{\lambda/2} {\delta \Psi_{,\theta \theta} \over \Psi^{\prime}_0} 
     - {1 \over 2 e^{\lambda/2}} \left(\delta g^{\prime}_{2 2} 
     - {r \over e^{\lambda}} \delta g_{1 1}\right) \ , \cr
     && \cr
     K_{3 3} &=& - {\rm sin}^2\theta \left({r \over e^{\lambda/2}} + 
     {\rm cot}\theta e^{\lambda/2} {\delta \Psi_{,\theta} \over 
     \Psi^{\prime}_0} + {1 \over 2 {\rm sin}^2\theta e^{\lambda/2}} 
     \delta g^{\prime}_{3 3} - {r \over 2 e^{3 \lambda/2}} \delta 
     g_{1 1}\right) \ ,
\end{eqnarray}
for the non-trivial components of the second fundamental form.  

Given that a smooth background can be constructed independently of the 
oscillations, we can assume that the background and linearized pieces of 
$\gamma_{\mu \nu}$ and $K_{\mu \nu}$ are separately continuous at the 
core-envelope interface.  We will consider the background pieces first.  
For clarity of presentation the matter and metric variables in the 
envelope will be distinguished by a tilde.  The continuity of the first 
fundamental form thus yields for the background
\beq
     \nu(R_c) = \tilde{\nu}(R_c) \ , 
\eeq
and continuity of the second fundamental form implies
\beq
     \nu^{\prime}(R_c) = \tilde{\nu}^{\prime}(R_c) \quad , \quad 
     e^{\lambda(R_c)} = e^{\tilde{\lambda}(R_c)} \ .
\eeq
Using the background Einstein equations, and defining
\beq
    e^{- \lambda} = 1 - 2 m(r)/r \quad , \quad 
    e^{- \tilde{\lambda}} = 1 - 2 \tilde{m}(r)/r \ , 
\eeq
where
\beq
    m(r) = - 4 \pi \int_0^r  r^2 \Lambda_0(r) dr \quad , \quad
    \tilde{m}(r) = - 4 \pi \int_{R_c}^r  r^2 \tilde{\Lambda}_0(r) dr 
                   + C \ ,
\eeq
we find that the background junction conditions imply $C = m(R_c)$ 
and $\Psi_0(R_c) = \tilde{\Psi}_0(R_c)$, where $\tilde{\Psi}_0$ and 
$- \tilde{\Lambda}_0$ are the pressure and energy density of the envelope, 
respectively.  It is also useful to note that the 
Tolman-Oppenheimer-Volkov equations for the core and envelope imply
\beq
    {\Psi^{\prime}_0(R_c) \over \tilde{\Psi}^{\prime}_0(R_c)} = 
    {\Psi_0(R_c) - \Lambda_0(R_c) \over \tilde{\Psi}_0(R_c) - 
    \tilde{\Lambda}_0(R_c)} \ . \label{pderivcont}
\eeq

Before dealing with the linear perturbations, it will be convenient to 
write out the linearized pressure as a function of the fundamental matter 
and metric variables, for both the core and the envelope.  We have 
used the field equations to help simplify the formulas.  The final forms 
are (for the radial dependence)
\beq
    \delta \Psi = \left[{\nu^{\prime} \over 2 r e^{\lambda/2}} \left(\mu 
                  \n W_\n + \chi p W_\p\right) - {1 \over e^{\nu/2}} 
                  \left(X_\n + X_\p\right)\right] r^l
\eeq
and
\beq
    \delta \tilde{\Psi} = \left[{\tilde{\nu}^{\prime} \over 2 r 
                          e^{\tilde{\lambda}/2}} \tilde{\chi} \tilde{\p} 
                          \tilde{W}_\p - {1 \over e^{\tilde{\nu}/2}} 
                          \tilde{X}_\p\right] r^l
\eeq
where we have again put a tilde over all the linearized variables 
associated with the envelope.  Now, it can be seen 
that the junction 
conditions imply that the metric perturbations are continuous at 
the core/envelope interface, i.e. 
\begin{eqnarray}
      H_0(R_c) &=& \tilde{H}_0(R_c) \ , \cr
               && \cr 
      H_1(R_c) &=& \tilde{H}_1(R_c) \ , \cr
               && \cr
      K(R_c) &=& \tilde{K}(R_c) \ .
\end{eqnarray}
The matter variables, on the other hand, must satisfy two conditions, 
which are
\beq
     \tilde{\chi}(R_c) \tilde{\p}(R_c) \tilde{W}_\p(R_c) = \mu(R_c) n(R_c) 
               W_\n(R_c) + \chi(R_c) \p(R_c) W_\p(R_c)  
\eeq
and
\begin{eqnarray}
     \tilde{X}_\p(R_c) &=& {\tilde{\Psi}_0(R_c) - \tilde{\Lambda}_0(R_c) 
               \over \Psi_0(R_I) - \Lambda_0(R_c)} \left(\Xn(R_c) + 
               \Xs(R_c)\right) - {\nu^{\prime}(R_c) e^{(\nu(R_c) - 
               \lambda(R_c))/2} \over 2 R_c} \left({\tilde{\Psi}_0(R_c) - 
               \tilde{\Lambda}_0(R_c) \over \Psi_0(R_c) - \Lambda_0(R_c)} 
               - 1\right) \times \cr 
               && \cr
               &&\tilde{\chi}(R_c) \tilde{\p}(R_c) \tilde{W}_\p(R_c) \ . 
               \label{xcont}
\end{eqnarray}

It is important to notice that 
the junction conditions do {\it not} imply 
that $\delta \Psi = \delta \tilde{\Psi}$ at the core/envelope interface, 
but rather that $\delta \Psi/\Psi_0^{\prime} = \delta \tilde{\Psi}/
\tilde{\Psi}_0^{\prime}$.  This fact is the clearest demonstration why a 
geometrical approach is crucial for obtaining the correct junction 
conditions, since this particular result is a direct consequence of the 
fact that Schwarzschild-like coordinates cause derivative discontinuities 
whenever the energy density is not continuous.  If it is the case that 
$\Lambda_0(R_c) = \tilde{\Lambda}_0(R_c)$, then we can see from 
Eq.~(\ref{pderivcont}) that the junction conditions {\it will} imply 
$\delta \Psi = \delta \tilde{\Psi}$.  Having a continuous energy density 
also implies that $\tilde{X}_\p(R_c) = X_\n(R_c) + X_\p(R_c)$.

\section*{Appendix II: The Analytical Equation of State}

In this Appendix we develop the analytical equation of state used in 
the main part of the text. The essential strategy is to introduce an
expansion based on the assumption that the fluid velocities
are small compared to the speed of light. This is a reasonable assumption 
for neutron star pulsations. 

\subsection{A Local Analysis of the Entrainment Parameter}

Recall that the entrainment variable $x^2$ is given by $x^2 = - \n^{\mu} 
\p_{\mu}$.  It is convenient to write each conserved four-current as in 
the main text: 
\beq
     \n^\rho = \n u^\rho \quad , \quad \p^\rho = \p v^\rho \ ,
\eeq
except that now we take $u^{\rho} u_{\rho} = - c^2$ and $v^{\rho} 
v_{\rho} = - c^2$ where $c$ is the speed of light.  If $\tau_{\n}$ and 
$\tau_{\p}$ denote the proper times of the neutron and proton fluid 
elements, respectively, then the worldlines of each fluid element are 
obtained from the functions
\beq
  x_{\n}^{\mu}(\tau_{\n}) = (t(\tau_{\n}),x_{\n}^i(\tau_{\n}))
        \quad , \quad
  x_{\p}^{\mu}(\tau_{\p}) = (t(\tau_{\p}),x_{\p}^i(\tau_{\p})) \ .
\eeq
The ``unit'' four-velocities are thus given by
\beq
    u^{\mu} = {dx_{\n}^\mu \over d\tau_{\n}} \quad , 
              \quad 
    v^{\mu} = {dx_{\n}^\mu \over d\tau_{\p}} \ .
\eeq
 
Consider, for the moment, a region within the fluid that is small 
enough that the gravitational field does not change appreciably across 
the region.  In this case, a locally Minkowski coordinate system can 
be set up and the metric can be approximated by the flat metric:
\beq
    ds^2 = - d(c t)^2 + \delta_{ij} dx^i dx^j 
                 \ .
\eeq
Letting $u_{\n}^i = dx_{\n}^i/dt$ and $v_{\p}^i = d
x_{\p}^i/dt$, as well as $u_{\n}^2 = \d_{ij}u_{\n}^i u_{\n}^j$ 
and $v_{\p}^2 = \d_{ij} v_{\p}^i v_{\p}^j$, one can show that the 
four-velocity components can be written as
\beq
    u^0 = {c \over \sqrt{1 -  (u_{\n}/c)^2}} \quad , \quad 
    u^i = {u_{\n}^i \over \sqrt{1 -  (u_{\n}/c)^2}} \ ,
\eeq
and
\beq
    v^0 = {c \over \sqrt{1 -  (v_{\p}/c)^2}} \quad , \quad 
    v^i = {v_{\p}^i \over \sqrt{1 -  (v_{\p}/c)^2}} \ .
\eeq

With this decomposition one finally obtains for the entrainment variable
\beq
    x^2 = \n \p \left({1 - \d_{i j} (u_{\n}^i/c) (v_{\p}^j/c) \over 
          \sqrt{1 - (u_{\n}/c)^2} \sqrt{1 - (v_{\p}/c)^2}}\right) \ .
\eeq
If it is the case that the individual three-velocities are small 
with respect to the speed of light, i.e.~that
\beq
     {\left|u_{\n}^i\right| \over c} << 1 \quad , \quad 
     {\left|v_{\p}^i\right| \over c} << 1 \ ,
\eeq
then it will be true that $x^2 \approx \n \p$ to leading order in the 
ratios $u_{\n}/c$ and $v_{\p}/c$.  This basic fact will be at the heart 
of the expansion considered below.

\subsection{The Analytical Equation of State}

Given what was just discussed, it makes sense to consider equations 
of state that can be expanded like
\beq
    \Lambda(n^2,p^2,x^2) = \sum_{i = 0}^{\infty} \lambda_i(n^2,p^2) 
                           \left(x^2 - \n \p\right)^i \ ,
\eeq
since $x^2 - \n \p$ is small with respect to $\n \p$.  In terms of this 
expansion, one can show the following for the $\A$, $\a00$, 
etc.~coefficients that appear in the field equations:
\begin{eqnarray}
  \A &=& - \sum_{i = 1}^{\infty} i~\lambda_i(n^2,p^2) \left(x^2 - \n 
         \p\right)^{i - 1}  \ , \cr
     && \cr
  \B &=& - {1 \over \n} {\partial \lambda_0 \over \partial \n} - {\p 
         \over \n} \A - {1 \over \n} \sum_{i = 1}^{\infty} {\partial 
         \lambda_i \over \partial \n} \left(x^2 - \n \p\right)^i \ , \cr
     && \cr
  \C &=& - {1 \over \p} {\partial \lambda_0 \over \partial \p} - {\n 
         \over \p} \A - {1 \over \p} \sum_{i = 1}^{\infty} {\partial 
         \lambda_i \over \partial \p} \left(x^2 - \n \p\right)^i \ , \cr
     && \cr
  \a00 &=& - {\partial^2 \lambda_0 \over \partial \p \partial \n} - 
           \sum_{i = 1}^{\infty} {\partial^2 \lambda_i \over \partial \p 
           \partial \n} \left(x^2 - \n \p\right)^i \ , \cr
     && \cr
  \b00 &=& - {\partial^2 \lambda_0 \over \partial \n^2} - 
           \sum_{i = 1}^{\infty} {\partial^2 \lambda_i \over \partial 
           \n^2} \left(x^2 - \n \p\right)^i \ , \cr
     && \cr
  \c00 &=& - {\partial^2 \lambda_0 \over \partial \p^2} - 
           \sum_{i = 1}^{\infty} {\partial^2 \lambda_i \over \partial 
           \p^2} \left(x^2 - \n \p\right)^i \ .  
\end{eqnarray}
It is important to note that $\a00$ vanishes if the master function 
is such that $\lambda_i$ are separable in $n$ and $p$.

The utility of this expansion is especially apparent for the 
quasinormal mode calculations, because when any of the coefficients are 
evaluated on the background, then one sets $x^2 = \n \p$, and thus only 
the first few $\lambda_i$ are needed. In fact, in our analysis we only retain
$\lambda_0$ and $\lambda_1$, where the latter contains the information 
concerning the entrainment effect. 
 
\section*{Appendix III: An accurate method for determining long-lived QNMs}

The problem of calculating quasinormal modes of relativistic systems, such
as black holes and neutron stars, is in many ways far from trivial.  In 
general, a strategy for finding QNMs has to involve a 
prescription for imposing a pure outgoing wave boundary condition at 
infinity for a linear second-order differential equation.  
In the context of the present paper the relevant 
equation is the one derived by Zerilli \cite{Z70}:
\beq
    \left[{d^2 \over dr_\ast^2} + \omega^2 - V(r) \right] Z  = 0 \ .
    \label{zerilli}
\eeq
Here $r_\ast$ is the standard tortoise coordinate and we have assumed 
that the perturbed quantities have a harmonic dependence on time, 
i.e. behave as $\exp(i \omega t)$.  The effective potential $V(r)$ 
is rather complicated but here we need only know that it
is such that  the behaviour of a general solution at spatial 
infinity (as $r_\ast \to +\infty$) is
\beq
    Z \sim A_{\rm out}(\omega) e^{-i\omega r_\ast} + A_{\rm in}(\omega) 
           e^{i\omega r_\ast} \ .
\eeq

A QNM of the system is a solution that combines some physical constraints 
(no waves coming out of the event horizon in the case of a black hole or 
a regular solution to the equations for the interior of a star) with 
purely outgoing waves, $A_{\rm in} (\omega_n)=0$,  at infinity.  Two 
typical difficulties arise in the determination of such mode solutions.  
Both are due to the fact that the QNMs are damped by gravitational 
radiation emission, and thus the QNM frequency $\omega_n$ must be complex 
(with a positive imaginary part unless the mode is unstable).   This means 
that an outgoing-wave solution to Eq.~(\ref{zerilli}) will be 
exponentially growing as $r_\ast \to +\infty$ and one would, in principle, 
need exponentially high numerical precision to discard the ingoing solution.  This 
problem is particularly challenging for rapidly damped modes, like those
of black holes and the neutron star w-modes.  A second difficulty that 
arises 
is relevant for the p-modes of a neutron stars.  The high order p-modes 
are damped very slowly by gravitational radiation.  Thus, the 
characteristic frequencies are such that the imaginary part is many orders 
of magnitude smaller than the real part.  Although not as conceptually 
challenging as the first QNM problem, the difficulty associated with 
extremely small imaginary parts typically prohibits the determination of 
any but the first few of the neutron star p-modes. In this Appendix, we 
describe a new method for dealing with this problem.  The  various mode results 
presented in the main body of the paper were obtained within this scheme. 

Let us assume that our relativistic system has a QNM with complex frequency
$\omega_n$.  Then the solution to the Zerilli equation is such 
that $A_{\rm in}(\omega_n) = 0$ as $r_\ast \to \infty$ whereas 
$A_{\rm out}(\omega_n) \ne 0$ asymptotically.  From the fact that it is 
the frequency squared that appears in all the relevant perturbation 
equations (see 
Sec.~\ref{eqns}) we can draw two general conclusions: i)  
another outgoing-wave mode is characterized by $-\bar{\omega}_n$ (where 
the bar represents complex conjugation), and ii) time-reversal of an 
outgoing-wave mode leads to a solution that corresponds to purely ingoing 
waves at infinity.  That is, a solution such that, at infinity, 
$A_{\rm out}(- \omega_n) = 0$ while $A_{\rm in}(-\omega_n) \ne 0$.  
Moreover, it follows that a second such solution corresponds to 
$\bar{\omega}_n$.  As we will demonstrate below, 
this information can be very useful when trying to 
identify normal-mode frequencies that have small imaginary parts.
 
\begin{figure}[h]
\centering
\includegraphics[height=7cm,clip]{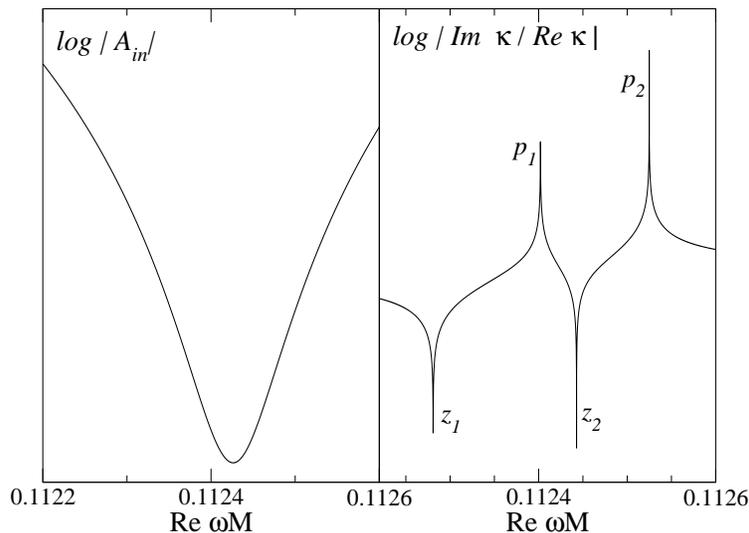} 
\caption{A graph of the incoming amplitude $A_{\rm in}$ vs the real part 
of the frequency.  In the left panel we see a standard ``Breit-Wigner 
resonance.'' The right panel illustrates the properties of the phase of the 
ration of the asymptotic amplitudes. We use 
the location of the two poles and zeros to deduce the frequency and damping 
rate of a long-lived pulsation mode.}
\label{ainfig}
\end{figure}

Most schemes for determining QNMs are based on numerical constructions of 
the ratio $\kappa = A_{\rm out}/A_{\rm in}$.  Assuming that
\beq
    A_{\rm in}(\omega) \sim (\omega -\omega_n)
\eeq
close to a QNM one can try to iterate for the zero of $A_{\rm in}$ using a 
standard scheme such as M\"uller's method.  This strategy works fine for 
rapidly damped modes (and typically also for the fluid f-mode), but does 
 usually not provide a reliable estimate for an imaginary part that is 
several orders of magnitude smaller than the real part.  An alternative 
method is inspired by resonant scattering problems in quantum theory.  
Letting $\omega_n = \alpha_n + i \beta_n$ (with $\alpha_n$ and $\beta_n$
both positive) we can easily identify the real part $\alpha_n$ from the 
position of the minimum of the standard ``Breit-Wigner resonance'' 
in a graph of $\log | A_{\rm in} |$, cf.~the example shown in 
Fig.~\ref{ainfig}.   It is not all that simple to extract the imaginary 
part, however.  To do this one must approximate the half-width of the 
peak, eg.~by curve fitting.  To achieve satisfactory precision in this 
process is not trivial.

To devise a better scheme for determining the damping rate of a very 
long-lived QNM, we employ the two general properties we deduced 
earlier.  From these it is clear that, if we have a zero of $A_{\rm in}$ 
close to the real frequency axis, there must also be a zero of 
$A_{\rm out}$ on the opposite side of the axis.  How does that alter the 
above results?  Close to a zero of $A_{\rm out}$ we will have
\beq
    A_{\rm out}(\omega) \sim (\omega -\bar{\omega}_n) \ .
\eeq
Then we immediately see that the absolute value of the ratio between the asymptotic 
amplitudes,  $\kappa$,  is a smooth, slowly varying function of the 
frequency.  But if we consider its phase, we find some interesting and 
useful features.  We get (for real $\omega$)
\beq
    \kappa \approx  \gamma {\omega - \alpha_n + i \beta_n \over \omega 
                     - \alpha_n - i \beta_n} 
\eeq 
where $\gamma=\gamma_r + i \gamma_i$ is a complex ``constant,'' and 
therefore
\beq
    {{\rm Im}\ \kappa \over {\rm Re}\ \kappa} \approx  {\gamma_i [(\omega 
    - \alpha_n)^2 - \beta_n^2] + 2 \gamma_r \beta_n (\omega-\alpha_n) 
    \over \gamma_r [(\omega - \alpha_n)^2 - \beta_n^2] - 2 \gamma_i 
    \beta_n (\omega - \alpha_n)} \ .
\eeq
From this we see that, instead of having a singularity at $\omega = 
\alpha_n$ we now have a function with two zeros and two poles on the 
real frequency axis, cf.~Fig.~\ref{ainfig}.  Provided that the calculation 
of the asymptotic amplitudes can be done with sufficiently 
high precision, the location 
of these zeros ($z_{1,2}$) and poles ($p_{1,2}$) can readily be deduced.  
Given this information one can show that
\begin{eqnarray}
    \alpha_n &\approx& {p_1 p_2 - z_1 z_2 \over p_1 + p_2 - z_1 -  z_2} \\
    \beta_n &\approx& [\alpha_n (z_1 + z_2) - \alpha_n^2 - z_1 z_2]^{1/2} 
    \\
    {\gamma_r \over \gamma_i} &\approx& {2 \beta_n \over p_1 + p_1 - 2 
    \alpha_n} = {1 \over 2 \beta_n} [2 \alpha_n - z_1 - z_2] \ .
\end{eqnarray}
It should be pointed out that one gets four equations for the real and 
imaginary parts of the QNM frequency, as well as the ratio $\gamma_r /
\gamma_i$.  The last equality above can therefore be used as a ``sanity 
check'' on the calculation.  In essence, it provides information about 
the accuracy of the obtained quantitites. 

In practise, this method works very well even when the imaginary part of 
the frequency is more than eight orders of magnitude smaller than the 
real part, cf.~the results in Table~\ref{freqtab2}.

\end{document}